\documentclass[useAMS,usenatbib]{mn2e}
\usepackage{amssymb}
\usepackage{epsfig}

\def\mnras{MNRAS}  
\def\apj{ApJ}      
\def\apjl{ApJL}    
\def\apjs{ApJS}    
\def\aap{A\&A}     
\def\araa{ARA\&A}  
\def\nat{Nature}   
\def\owls{{\small OWLS}}
\def\pasp{PASP}

\def\igm{IGrM}

\title[How quasar winds solve over-cooling in groups]{Gas expulsion by quasar-driven winds as a solution to the over-cooling problem in galaxy groups and clusters}
\author[I. G. McCarthy et al.]{I. G. McCarthy$^{1,2,3}$\thanks{E-mail:
mccarthy@ast.cam.ac.uk (IGM)}, J. Schaye$^4$, R. G. Bower$^5$, T. J. Ponman$^6$, C. M. Booth$^4$, \newauthor C. Dalla Vecchia$^{4,7}$, V. Springel$^{8,9}$
\\
\\
$^{1}$Kavli Institute for Cosmology, University of Cambridge, Madingley Road, Cambridge, CB3 OHA\\
$^{2}$Astrophysics Group, Cavendish Laboratory, JJ Thomson Avenue, Cambridge, CB3 0HE\\
$^{3}$Institute of Astronomy, University of Cambridge, Madingley Road, Cambridge, CB3 0HA\\ 
$^{4}$Leiden Observatory, Leiden University, P. O. Box 9513, 2300 RA Leiden, the Netherlands\\
$^{5}$Institute of Computational Cosmology, Department of Physics, University of Durham, Science Laboratories, South Road, Durham DH1 3LE\\
$^{6}$Astrophysics and Space Research Group, School of Physics and Astronomy, University of Birmingham, Edgbaston, Birmingham, B15 2TT\\
$^7$Max Planck Institute for Extraterrestrial Physics, Giessenbachstrabe 1, 85748 Garching, Germany\\
$^8$Heidelberg Institute for Theoretical Studies, Schloss-Wolfsbrunnenweg 35, 69118 Heidelberg, Germany\\
$^9$Centre for Astronomy, Heidelberg University, M\"{o}nchhofstr. 12-14, 69120 Heidelberg,Germany
}

\begin{document}

\date{Accepted XXXX. Received XXXX; in original form XXXX}

\pagerange{\pageref{firstpage}--\pageref{lastpage}} \pubyear{2010}

\maketitle

\label{firstpage}

\begin{abstract}

Galaxy groups are not scaled down versions of massive galaxy
clusters - the hot gas in groups (known as the intragroup medium, \igm\
for short) is, on average, less dense than the intracluster medium,
implying that one or more non-gravitational processes (e.g., radiative
cooling, star formation, and/or feedback) has had a relatively larger
effect on groups. In the present study, we compare a number of
cosmological hydrodynamic simulations that form part of the
OverWhelmingly Large Simulations project to isolate and quantify the
effects of cooling and feedback from supernovae (SNe) and active
galactic nuclei (AGN) on the gas. This is achieved by comparing
Lagrangian thermal histories of the gas in the different runs, which
were all started from identical initial conditions. While radiative
cooling, star formation, and SN feedback are all necessary ingredients,
only runs that also include AGN feedback are able to successfully
reproduce the optical and X-ray properties of groups and low-mass clusters. We isolate how,
when, and exactly what gas is heated by AGN. Interestingly, we find that
the gas that constitutes the present-day \igm\ is that which was {\it
not} strongly heated by AGN. Instead, the low median density/high median entropy of 
the gas
in present-day groups is achieved by the ejection of lower entropy gas
from low-mass progenitor galaxies at high redshift (primarily $2 \la z
\la 4$). This corresponds to the epoch when supermassive black holes
accreted most of their mass, typically at a rate that is close to the
Eddington limit (i.e., when the black holes are in a `quasar mode').

\end{abstract}


\section{Introduction}

Understanding the formation and evolution of galaxy groups is crucial to solving the general problem of galaxy formation, as groups contain most of the galaxies in the universe at the present day (e.g., Mulchaey 2000), and a large fraction of the universal stellar mass is formed in groups (e.g., Crain et al.\ 2009).  They can also potentially yield important insights into the formation and evolution of more massive and rare galaxy clusters, since in the current cosmological paradigm clusters acquire most of their mass through the accretion/mergers of groups.  Groups of galaxies are special systems in an observational sense as well, as it is possible to probe both the stellar and hot gas components of these systems directly, whereas for normal galaxies detecting the diffuse hot gas is presently very challenging.  As highlighted in many recent theoretical papers (e.g., Bower et al.\ 2008, hereafter B08; McCarthy et al.\ 2010; Fabjan et al.\ 2010; Puchwein et al.\ 2010), the properties of the hot gas and of the stellar component are intimately linked and any successful model must be able to reproduce both simultaneously.

X-ray observations of nearby, relatively bright galaxy groups have revealed that the hot gas (the intragroup medium, \igm\ for short) has some puzzling properties.  In particular, it is on average less dense than the hot gas found in massive clusters (known as the intracluster medium, ICM for short).  This is not what one expects if gravitational processes alone determine the thermodynamic properties of the gas (e.g., Voit, Kay, Bryan 2005, hereafter VKB05).  The observed decrease of the gas density with halo mass therefore indicates that one or more non-gravitational processes (e.g., radiative cooling, star formation, and/or feedback) has had a larger effect on groups than on massive clusters.  The fact that a non-negligible fraction of the total mass of baryons in groups is locked up in stars (e.g., Balogh et al.\ 2001; Lin \& Mohr 2004; Gonzalez et al.\ 2007; Giodini et al.\ 2009) is already testament that non-gravitational processes are important on the scale of groups.

The entropy\footnote{As is common in X-ray astronomy, we define the `entropy', $S$, as $k_B T/n_e^{2/3}$ and use units of keV cm$^2$, where $k_B$ is Boltzmann's constant, $T$ is the temperature of the gas and $n_e$ is the electron number density.  The observational entropy $S$ is related to the thermodynamic specific entropy $s$ via $s \propto \ln{S}$ and, like the thermodynamic entropy, $S$ will be conserved in any adiabatic process.} of the \igm\ is perhaps the best tool for studying the impact of non-gravitational physics, since it maintains a record of the thermal history of the gas  (see Voit et al.\ 2003 and Voit 2005 for further discussion).  Using {\it ROSAT} observations, Ponman et al.\ (1999) were the first to show convincingly that galaxy groups have `excess' entropy relative to massive clusters; i.e., when one scales the radius and entropy by some suitable characteristic radius and entropy that depend only on system mass (e.g., the virial radius and `virial entropy'; see Section 3.1), groups typically have higher scaled entropies at all scaled radii than clusters. And both groups and clusters have excess entropy relative to theoretical models that only account for gravitational heating (e.g., Balogh et al.\ 1999; Babul et al.\ 2002; Voit et al.\ 2002).  More recent observations, particularly with {\it ASCA}, {\it Chandra}, and {\it XMM-Newton}, have improved the quality of the measurements and extended both the range of the masses and radii over which the entropy has been measured (e.g., Finoguenov et al.\ 2002; Ponman et al.\ 2003; Sun et al.\ 2009; Johnson et al.\ 2009; Pratt et al.\ 2010; Giodini et al.\ 2010).  These newer results have generally confirmed those from the {\it ROSAT} satellite: groups contain excess entropy relative to clusters as well as to models that only take into account gravitational heating.

On the theory side, many suggestions have been put forward for the physical origin of the excess entropy.  The most commonly mentioned are `preheating' (first introduced by Kaiser 1991 and Evrard \& Henry 1991), radiative cooling (e.g., Bryan 2000), and supernova (SN) and active galactic nuclei (AGN) feedback within groups.  Preheating is conceptually simple: at early times the entropy of the {\it proto}-\igm\ is raised by feedback processes (such as outflows driven by SN or AGN).  The increased entropy prevents the gas from being easily accreted or compressed, resulting in a lower mean density (and gas mass fraction) for groups.  Note that for a fixed level of entropy injection the effect is larger for groups than for clusters, since groups have lower entropies than clusters.  Somewhat paradoxically, radiative cooling has also been shown to raise the entropy of the gas in groups and clusters.  It does so not by heating the gas, but by selectively removing the lowest entropy gas from the hot \igm\ phase, turning it into cold gas and stars, and thereby raising the mean entropy of the hot gas that is unable to cool (e.g., Bryan 2000).  Feedback from winds driven by SNe and jets/winds launched by supermassive black holes (hereafter BHs) originating from orbiting group galaxies may also significantly raise the entropy of the \igm\ by direct heating.  At present, no clear consensus has emerged as to which of these processes are most important for raising the entropy of the gas over and above that generated through gravitational shock heating alone.

In the present study, we seek to develop a better understanding of the physical processes that set the {\it global} thermodynamic properties of the IGrM.  We will do this by exploiting a suite of high resolution cosmological hydrodynamical simulations which form part of the OverWhelmingly Large Simulations project ({\small OWLS}; Schaye et al.\ 2010).  The purpose of the \owls\ project is to explore the effects of varying so-called `subgrid' physics in the simulations, such as radiative cooling, star formation, and SN and BH feedback.  All simulations are started from identical initial conditions, with each simulation varying a parameter within one of the subgrid physics modules or switching on/off one of the subgrid processes entirely.  By comparing the different simulations, it is possible to isolate which processes are most important for raising the entropy of the IGrM.  In particular, we will compare not only the present-day entropy distributions of the groups in the different simulations, but also, for the first time, the {\it Lagrangian histories} of the gas in the different runs.  This allows us to follow the heating and cooling of a parcel of gas over cosmic time.  We will show that a particularly powerful discriminator is to identify and follow the thermal properties of the same sets of particles in the (identical) initial conditions of the various simulations.  

In McCarthy et al.\ (2010; hereafter Paper I) we showed that the \owls\ {\it AGN} model yields an excellent match to a wide range of optical and X-ray observations of nearby galaxy groups, including the entropy, temperature, and metallicity distribution and their dependence on halo mass, as well as the stellar mass of the group and the central brightest galaxy (CBG) and the stellar age of the CBG.  Based on comparisons between this model and a number of other models, we argue below that, in fact, {\it none} of the three physical scenarios outlined above provides an accurate description for how the entropy of the \igm\ was raised.  Instead, we show that gas that constitutes the present-day \igm\ was that which {\it escaped} significant AGN feedback and the excess entropy is due primarily to the ejection of the lowest entropy gas at high redshift (i.e., the gas that was significantly heated is no longer within the virialized region).

The present paper is organised as follows.  In Section 2 we provide a brief description of the simulations.  In Section 3 we present our main results on the thermal history of the \igm\ in the simulations.  In Section 4 we explore further how the AGN feedback raises the entropy of the IGrM.  Finally, in Section 5 we summarize and discuss our findings.  All quantities presented in this study (e.g., entropy, temperature, density) are in physical, rather than co-moving, units.

\section{Simulations}

An in-depth presentation of the \owls\ project can be found in Schaye et al.\ 
(2010).  We therefore present only a brief description of the simulations here.

All of the \owls\ runs we analyse are initialised from identical initial conditions, in particular from a $\Lambda$CDM cosmological density field (with a power spectrum computed using CMBFAST) in a periodic box of $100 h^{-1}$ Mpc on a side at $z=127$.  In generating this density field, the various relevant cosmological 
parameters ($\Omega_b$, $\Omega_m$, $\Omega_\Lambda$, $h$, $\sigma_8$, $n_s$) were set 
equal to the maximum-likelihood values found from the analysis of the 3-year {\it WMAP} 
cosmic microwave background data (0.0418, 0.238, 0.762, 0.73, 0.74, 0.95, respectively; Spergel et al.\ 2007).  The simulations are evolved to $z=0$ using the 
TreePM-SPH code {\small GADGET-3} (last described in Springel 2005).  We extract from the final snapshot ($z=0$) all galaxy 
groups for which $\log_{10}($M$_{200}/$M$_\odot) \ge 13.00$ (where M$_{200}$ is the total mass within 
a radius that encloses a mean density of 200 times the present-day critical density of 
the universe).  The simulations use $2\times512^3$ particles, yielding a particle mass for gas and dark matter of $m_{\rm gas} 
\approx 8.65 \times 10^7 h^{-1}$ M$_\odot$ and  m$_{\rm dm} \approx 4.06 \times 10^8 h^{-1}$ 
M$_\odot$, respectively. Thus, a $\sim 10^{14} h^{-1}$ M$_\odot$ galaxy group is 
resolved with $\sim 10^5$ gas and dark matter particles.  As shown in the Appendix of Paper I, this resolution is sufficient to robustly predict the properties of galaxy groups in the simulations.  

\begin{table*}
\caption{A list of all simulations, the corresponding names in the
  \owls\ project (Schaye et al.\ 2010) and the differences in the included
  subgrid physics.}
\centering
\begin{tabular}{llcccc} \hline
Simulation & OWLS name & Cooling & Star formation & Supernova & AGN \\
 & & & & Feedback & Feedback \\
\hline
{\it NoCool} & ... & None & No & No & No \\
{\it PrimCool\_SF} & NOSN-NOZCOOL & Primordial & Yes & No & No \\
{\it ZCool\_SF} & NOSN & Metal & Yes & No & No \\
{\it PrimCool\_SF\_SN} & NOZCOOL & Primordial & Yes & Yes & No \\
{\it ZCool\_SF\_SN} & REF & Metal & Yes & Yes & No \\
{\it ZCool\_SF\_SN\_AGN} & AGN & Metal & Yes & Yes & Yes \\
\hline
\end{tabular}
\label{tab:simlist}
\end{table*} 

In Paper I (see also Schaye et al.\ 2010), we described in some detail the implementation of `subgrid' physics (e.g., radiative cooling, star formation, and SN and AGN feedback) in the {\small GADGET-3} code and how we select and analyse galaxy groups.  We refer the reader to that study for details.  As discussed in Section 1, the thermal properties of the \igm\ differ from those predicted by simulations that include only gravity and basic non-radiative hydrodynamics.  As a result, it is now widely believed that one or more non-gravitational processes are responsible for modifying the properties of the hot gas.  The most likely culprits are cooling, star formation, and feedback, which are implemented in the \owls\ simulations using subgrid prescriptions.  The power of the \owls\ project is not only that it systematically varies the relevant physical parameters of a given subgrid physics module, but that it also systematically varies which subgrid processes are switched on/off.  This is the best and perhaps only way to isolate which processes are most important for establishing the thermodynamics properties of the IGrM.  

We note that the physics that raises the entropy of the ICM in massive clusters is quite likely to be closely related to that which operates on the scales of groups.  However, the volume of the \owls\ runs [($100 h^{-1}$ Mpc)$^3$] is not sufficiently large to contain massive galaxy clusters.  We have plans to carry out both `zoomed' initial condition and larger periodic box simulations in the future.

Below, we provide brief descriptions of the runs that we analyse in the present study.  The runs we have selected represent a range of interesting limiting cases and not the entire \owls\ suite of simulations.  In Table 1 we list all the simulations, the corresponding names in the \owls\ project and the differences in the included subgrid physics.

\subsection{Individual runs}

\begin{itemize}
\item{{\it NoCool}: This model omits the effects of radiative cooling, star formation, and feedback from SNe and BHs.  It does, however, include {\it net} photoheating from a Haardt \& Madau (2001) UV/X-ray background (i.e., whenever $\Lambda_{\rm net} \equiv \Lambda_{\rm heat}-\Lambda_{\rm cool} > 0$, otherwise $\Lambda_{\rm net} = 0$).  However, this photoheating has negligible consequences for the gas that constitutes the \igm\ at the present-day for redshifts of $z \la 2$ (see the Appendix). Therefore, this model is basically equivalent to a pure `non-radiative' run below $z \approx 2$.
We include net photoheating purely for consistency, since all other \owls\ runs include UV photoheating.  {\it This model will serve as our baseline for quantifying how radiative cooling, star formation, and feedback from SNe and AGN raise the entropy of the \igm\ over and above that of gravitational shock heating.}}

\item{{\it PrimCool\_SF}: This model is identical to the {\it ZCool\_SF} model (described below), except that cooling rates are calculated assuming primordial abundances.}

\item{{\it ZCool\_SF}: Radiative cooling rates for the gas are computed on an element-by-element basis by interpolating within pre-computed tables (generated with the CLOUDY software package, last described in Ferland et al.\ 1998) that contain 
cooling rates as a function of density, temperature, and redshift calculated in the 
presence of the cosmic microwave background and the Haardt \& Madau 
(2001) UV/X-ray background (see Wiersma et al.\ 2009a).  Star formation is 
tracked in the simulations following the pressure-based prescription of Schaye \& Dalla Vecchia (2008), which reproduces the observed Kennicutt-Schmidt law by construction.   
The timed release of individual elements by both massive (Type II SNe and 
stellar winds) and intermediate mass stars (Type Ia SNe and asymptotic giant branch stars) 
is included following the prescription of Wiersma et al.\ (2009b).  A set of 11
individual elements are followed in these simulations (H, He, C, N, O, Ne, Mg, Si, S, Ca, Fe), which represent all the important species for computing radiative cooling rates (Wiersma et al.\ 2009a).   As is the case for other \owls\ runs, UV/X-ray  photoheating is included. However, as already noted, this is unimportant below $z \approx 2$ and this model therefore effectively represents a `no feedback' model for galaxy groups.  (Note that even though there is no form of effective feedback in this run, metals are still removed from galaxies and dispersed into the \igm\ by ram pressure stripping.)}

\item{{\it PrimCool\_SF\_SN}: This model is identical to the {\it PrimCool\_SF} model except that feedback from SN-driven winds is also included.  In particular, the kinetic wind model 
of Dalla Vecchia \& Schaye (2008) is adopted with the wind velocity, $v_w$, set to $600$ km/s and the 
mass-loading parameter (i.e., the ratio of the mass of gas given a velocity kick to that 
turned into newly formed star particles), $\eta$, set to $2$.  This choice of parameters corresponds to using approximately 40\% of the total energy available SNe, for a Chabrier (2003) stellar initial mass function (IMF) 
(which is adopted in all of the simulations presented in this paper, with the obvious exception of {\it NoCool}).
}

\item{{\it ZCool\_SF\_SN}:  This model is identical to {\it PrimCool\_SF\_SN} except cooling rates are calculated on an element-by-element basis, as described above for the {\it ZCool\_SF} model.  {\it ZCool\_SF\_SN} was referred to as {\it REF} in Paper I (see Table 1).}

\item{{\it ZCool\_SF\_SN\_AGN}: This model is identical to the {\it ZCool\_SF\_SN} model except that it also includes the prescription for BH growth and AGN feedback of Booth \& Schaye (2009), which is a modified version of the algorithm developed by Springel et al.\ (2005).  A brief description is as follows.  BHs grow in mass through mergers with other BHs and through the accretion of neighbouring gas.  Gas is assumed to accrete at the local
Bondi-Hoyle-Lyttleton rate. However, if the local gas density exceeds
$n_{\rm H} = 0.1\,{\rm cm}^{-3}$ then this rate is multiplied by the
factor $[n_{\rm H} / (0.1\,{\rm cm}^{-2})]^2$ to compensate for the fact
that our simulations lack both the resolution and the physics to model a
cold, interstellar gas phase (see Booth \& Schaye 2009 for a discussion).  A certain fraction of the rest mass energy of the accreted gas, $\epsilon$, is used to 
heat a number of randomly selected neighbouring gas particles, $n_{\rm heat}$, by raising their temperatures by an amount $\Delta T_{\rm heat}$.  (Note that the BHs store feedback energy until it is sufficient to heat $n_{\rm heat}$ particles by $\Delta T_{\rm heat}$.)  This efficiency is a free parameter of the model.  Booth \& Schaye (2009) find that a value of $\epsilon=0.015$ yields a good match to the observed $z=0$ relations between BH mass and stellar mass and velocity dispersion and the $z=0$ cosmic BH density.  We therefore fix $\epsilon$ to this value.  As discussed below (see Section 4.4), the predicted properties of galaxy groups are insensitive to the adopted value of $n_{\rm heat}$ (which, for the present study, we set to unity) but {\it are} sensitive to $\Delta T_{\rm heat}$ if this quantity is set to a value that is comparable to (or lower than) the virial temperature of the high redshift progenitors of galaxy groups.  For specificity, we set $\Delta T_{\rm heat} = 10^8$ K and $n_{\rm heat} = 1$} .

\end{itemize}

To give a visual example of the importance of including non-gravitational processes in simulations of the \igm\ we present Fig.\ 1, which shows projected mass-weighted entropy, surface gas mass density, and projected mass-weighted temperature of the gas in and around the same randomly-selected galaxy group in {\it NoCool}, {\it ZCool\_SF\_SN}, and {\it ZCool\_SF\_SN\_AGN}.  The maps are $10 \ h^{-1}$ Mpc on a side.  Particles were interpolated to a 3D uniform grid (with $1000^3$ pixels) using the SPH smoothing kernel and projected along the z-axis of the simulation box.  In {\it NoCool} the hot, high entropy gas is confined to the area in the immediate vicinity of the group and to a large filament that runs from the bottom left to the top right of the maps.  The introduction of supernova feedback (and cooling and star formation; {\it ZCool\_SF\_SN}) results in a relatively small amount of moderate-entropy gas being expelled into voids.  The inclusion of AGN feedback has a much more dramatic effect, with large amounts of high entropy/high temperature gas being ejected out into the surrounding intergalactic medium.  We discuss in detail the importance of gas expulsion by AGN in Sections 3 and 4.

\section{The Thermal History of the IGrM}

In this section we first present the radial entropy distributions of the various \owls\ runs, demonstrating that the inclusion of cooling and feedback processes indeed raises the entropy of the hot gas in groups.  We then examine the entropy histories of these runs in detail and compare them with one another to deduce which processes are most important for raising the entropy of the gas.

\begin{figure*}
\includegraphics[width=18cm]{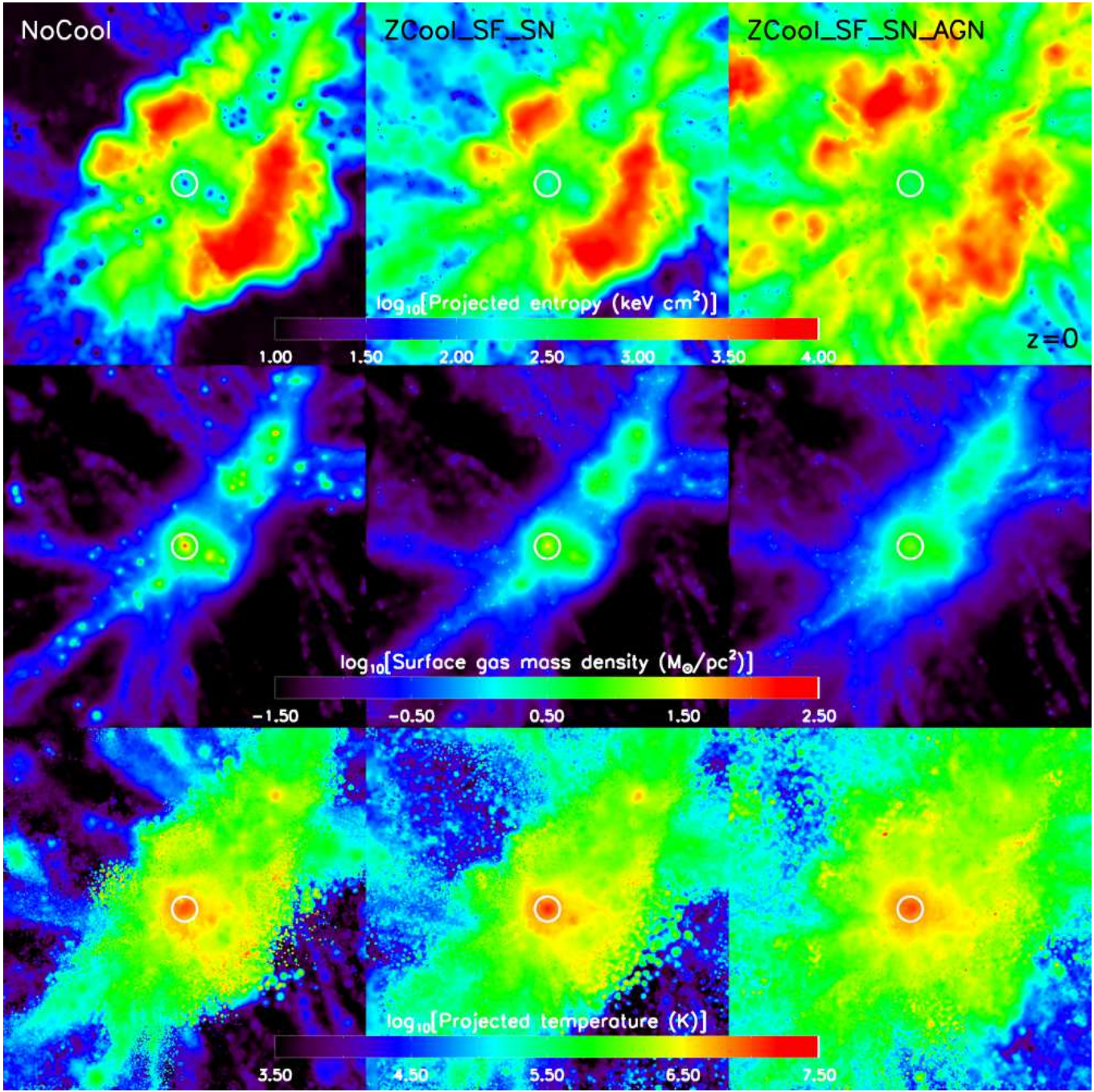}
\caption{Projected mass-weighted entropy (top), surface gas mass density (middle), and projected mass-weighted temperature (bottom) maps of a randomly-selected galaxy group at $z=0$ in three different runs {\it NoCool} (left), {\it ZCool\_SF\_SN} (middle), and {\it ZCool\_SF\_SN\_AGN} (right).  Each map is $10 \ h^{-1}$ Mpc on a side and centered on the group, which has a total mass of $M_{500} \approx 3.3\times10^{13} M_\odot$.  The solid white circle delineates $r_{500}$.  Feedback by AGN has ejected high entropy gas from the high-z progenitors of the group.
}
\label{fig:maps}
\end{figure*}

\subsection{Radial distribution of the entropy}

We begin by examining the 3D radial entropy distributions of the hot gas ($T>10^5$ K) in the \owls\ runs described in Section 2.1.  In Fig.\ 2 we plot the median 3D mass-weighted entropy profiles for all groups with masses between $13.25 \le \log_{10}(M_{500}/M_\odot) \le 14.25$ [approximately 70 groups with a median mass of $\log_{10}(M_{500}/M_\odot) \approx 13.5$], where $M_{500}$ is the total mass within 
the radius that encloses a mean density of 500 times the present-day critical density of the universe. The radial coordinate has been normalised by $r_{500}$, which is the radius that encloses M$_{500}$ (typically $r_{500} \approx 0.65 r_{200}$).  The entropy has been normalised by $S_{500}$, the `virial entropy', which we define, following VKB05, as:
\begin{eqnarray}
S_{500}(z=0) & \equiv & \frac{k_B T_{500}}{[n_{e,500}]^{2/3}} \\
& = & \frac{G M_{500} \mu m_H}{2 r_{500} [500 f_b \rho_{crit}(z=0)
/ (\mu_e m_H)]^{2/3}} \nonumber \\
& \propto & \biggl(\frac{M_{500}}{f_b}\biggr)^{2/3} \nonumber
\end{eqnarray}  

\noindent where $f_b \equiv \Omega_b/\Omega_m$, and $\mu_e$ is the mean molecular weight per free electron.  Note that $S_{500}$ is {\it not} the entropy at $r_{500}$, i.e., $S_{500} \ne S(r_{500})$, even in the case of the self-similar model.  This is because $S_{500}$ is defined in terms of the mean gas density within $r_{500}$ (as opposed to the gas density at $r_{500}$) for a halo with the universal baryon fraction and the virial temperature, $T_{500}$, which is not equivalent to the gas temperature at $r_{500}$ for non-isothermal mass distributions [for an NFW distribution, $T(r_{500}) \sim 0.6 T_{500}$, assuming hydrostatic equilibrium and that gas traces dark matter].  Therefore, $S_{500}$ depends only on the halo mass, which is dominated by dark matter, and is insensitive to the thermodynamic state of the gas. 

For reference we show in Fig.\ 2 the 16th and 84th percentiles of the {\it Chandra} entropy profiles of Sun et al.\ (2009) [approximately 20 groups with a median mass of 
$\log_{10}(M_{500}/M_\odot) \approx 13.5$; note we have used here only the 20 lowest mass groups of Sun et al.\ so as to achieve a similar median mass to that of the simulated groups].  Also shown (dotted line) is the power-law fit 
of VKB05 to the entropy profiles of a sample of groups and 
clusters simulated with non-radiative physics (see below).  

\begin{figure}
\includegraphics[width=\columnwidth]{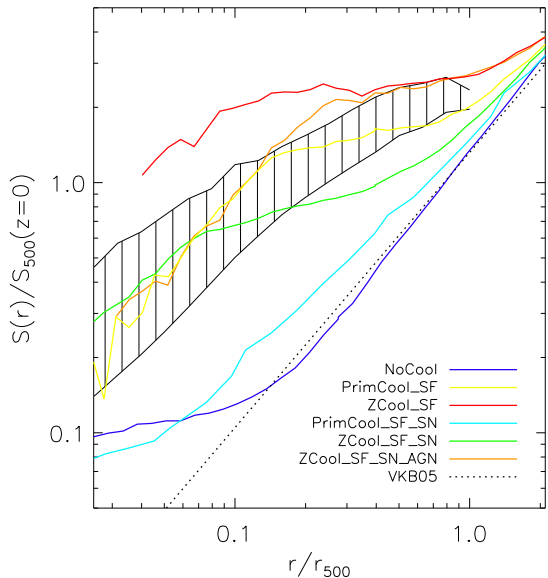}
\caption{Median 3D mass-weighted entropy profiles for a selection of \owls\ runs for groups with masses between $13.25 \le \log_{10}(M_{500}/M_\odot) \le 14.25$.  All other \owls\ runs not shown (see Schaye et al.\ 2010) fall in between {\it NoCool} and {\it ZCool\_SF}, in terms of their scaled entropy at intermediate radii.  For reference, the hatched region represents the observational data of Sun et al.\ (2009) and the dotted line is the power-law fit of VKB05 to $r > 0.3 r_{500}$ for a suite of non-radiative cluster simulations.  Note that at intermediate radii the runs without SN feedback (i.e., {\it ZCool\_SF} and {\it PrimCool\_SF}) yield {\it higher} entropy than the runs that include SN feedback, while the run that includes AGN feedback ({\it ZCool\_SF\_SN\_AGN}) yields a similar entropy as the run with metal-line cooling and without feedback ({\it ZCool\_SF}).}
\label{fig:entropy_profiles}
\end{figure}

Since we are interested primarily in investigating what sets the {\it global} thermodynamic properties of the IGrM, we will focus our attention on quantities such as the median gas particle entropy of groups.  Typically, the median entropy corresponds to the entropy at intermediate radii of $\sim0.4-0.6 r_{500}$.  From Fig.\ 2 we see that over this range in radii the simulations show a wide range of entropy values, spanning nearly an order of magnitude in $S/S_{500}$.  Of course, not all of the models we have included are physically reasonable, but several of them represent limiting cases which can be used to isolate the importance of certain subgrid physical processes, and the large differences between the models make it easier to do so.  

For example, {\it NoCool}, which is effectively a non-radiative run (note the agreement between the {\it NoCool} run and the power-law fit of VKB05, who fit approximately to the range $0.3-1.5 r_{500}$), yields the lowest entropy of any of the runs.  This is not unexpected, since the only process by which the entropy may be raised in this model is through gravitational shock heating.  Since gravity has no preferred scale, the resulting shape and normalisation of the entropy profile does not depend on system mass, which is why it is sometimes referred to as the `self-similar' profile.  Note that within $\approx 0.2 r_{500}$ the entropy profile deviates from a pure power-law behaviour, flattening into a core.  This was also found by VKB05 (see also Frenk et al.\ 1999) for both adaptive mesh refinement (AMR) and SPH non-radiative cosmological simulations.  The physical origin of the core appears to be related to the efficiency of energy exchange between gas and dark matter (McCarthy et al.\ 2007; see also Mitchell et al.\ 2009).

{\it ZCool\_SF} represents another interesting limiting case, as it includes cooling via both thermal bremsstrahlung and (metal) lines but no feedback.  This model yields very high entropies at intermediate radii.  That simulations {\it without} feedback can yield entropies in excess of the self-similar model has been noted previously (e.g., Muanwong et al.\ 2002; Dav{\'e} et al.\ 2002) and explained, at least qualitatively, by Knight \& Ponman (1997) and by Bryan (2000), Voit \& Bryan (2001), and Voit et al.\ (2002) (hereafter collectively referred to as VB).  These authors argued that radiative cooling acts to selectively remove the lowest entropy gas from the hot phase, turning it into cold gas and stars.  In the language of VB, radiative cooling `truncates' the entropy distribution of the gas (Voit \& Bryan 2001 and Voit et al.\ 2002 also noted the possibility that feedback could truncate the entropy distribution, by removing the lowest entropy gas).  As a result of this truncation, the mean entropy of the remaining hot gas increases with time\footnote{The entropy of an individual parcel of gas does not increase as a result of cooling (it decreases), but the mean/median entropy of the overall hot phase increases as a result of the removal of the lowest entropy gas from the hot phase over time due to cooling.}.  This hot gas, which was initially at large radii, due to its buoyancy, would then flow inwards, approximately adiabatically, to fill the pressure decrement left by the gas that was converted into cold (pressure-less) form.  As Fig.\ 2 emphasises, this results in a large amount of excess entropy in the remaining gas.  We will show below that this is because of the unrealistically large fraction of baryons that are converted into stars in this run.

The results of {\it PrimCool\_SF}, which is identical to {\it ZCool\_SF} except that the cooling rates are calculated assuming primordial abundances, are also consistent with the VB truncation hypothesis.  This run also yields entropy in excess of the {\it NoCool} (self-similar) result, but not to the same degree as {\it ZCool\_SF}.  This can be understood in terms of the reduced efficiency of cooling due to the neglect of metal-line emission in {\it PrimCool\_SF}.  The reduced cooling rate results in a smaller truncation of the entropy distribution of the hot gas which, in turn, leads to a smaller rise in the entropy.

The relatively large difference in the resulting entropy distributions of {\it ZCool\_SF} and {\it PrimCool\_SF} highlights the importance of metal-line cooling in the simulations.  Models without metal-line cooling will seriously underestimate the fraction of baryons in stars relative to models that do include metal-line cooling.  Comparisons of such models to observations can therefore quite easily lead to spurious conclusions about the required efficiency of feedback processes.  Note also that the difference in the resulting entropy distributions between runs with and without metal-line cooling is even larger when feedback is included (compare {\it PrimCool\_SF\_SN} and {\it ZCool\_SF\_SN}), since more particles are enriched due to expulsion of metals from star-forming regions by SNe, particularly at high redshift.
Unlike the \owls\ runs, most simulations presented in the literature calculate cooling rates assuming primordial abundances only or, if they do include metals, they do so by assuming abundances that are a fixed fraction of solar.

\begin{figure}
\includegraphics[width=\columnwidth]{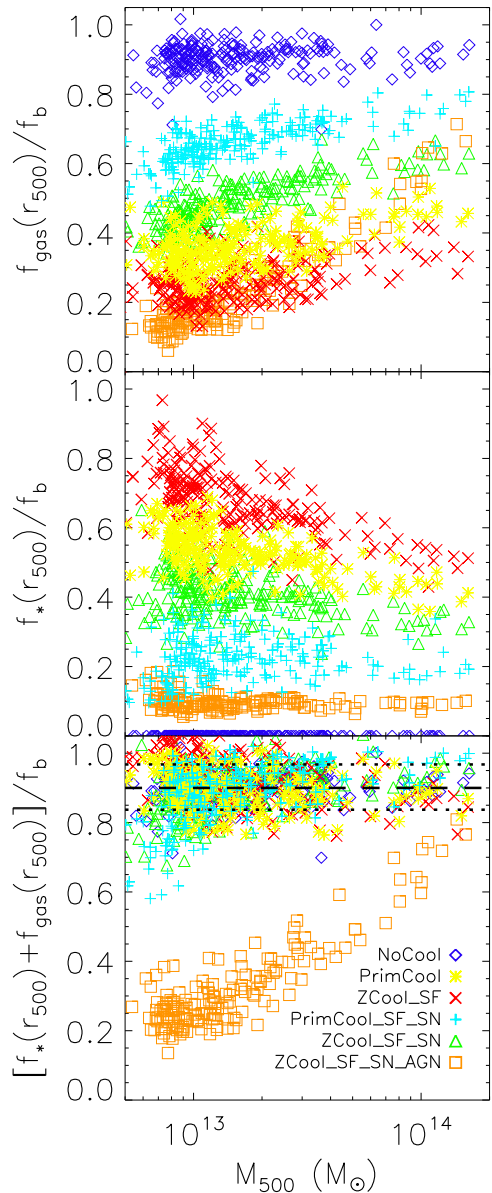}
\caption{Gas mass (top), stellar mass (middle), and total baryon (bottom) mass fractions within $r_{500}$ as a function of group mass, M$_{500}$, for the \owls\ runs investigated in this study.  The mass fractions have been normalised to the universal baryon fraction, $f_b = \Omega_b/\Omega_m$.  The dashed line in the bottom panel represents the median baryon fraction of the high mass systems in the SPH non-radiative Millennium Gas Simulation (Crain et al.\ 2007), while the dotted lines represent the 10th and 90th percentiles from that study.  Variations in the included subgrid physics can have a large effect on the predicted stellar and gas mass fractions but the total baryonic fraction is always near the universal value.  The one exception is the run that includes AGN feedback ({\it ZCool\_SF\_SN\_AGN}), for which the total baryon fraction is a steep function of halo mass.  Energy injection from BHs is able to eject gas from groups, particularly at high redshift. }
\label{fig:baryon_fractions}
\end{figure}

The other runs plotted in Fig.\ 2 ({\it ZCool\_SF\_SN}, {\it PrimCool\_SF\_SN}, {\it ZCool\_SF\_SN\_AGN}) also include feedback processes, so determining the relative importance of cooling and feedback in producing the excess entropy is not possible based on Fig.\ 2 alone.  Furthermore, we have not yet explicitly demonstrated that it is indeed the VB truncation mechanism that is responsible for the excess entropy of the hot gas in the models without feedback ({\it ZCool\_SF} and {\it PrimCool\_SF}).  In the remainder of this section we will exploit the Lagrangian nature of the \owls\ simulations to deduce which processes are most important for raising the entropy of the \igm\ in the simulations.  Of particular interest will be isolating the effects the AGN feedback, since only {\it ZCool\_SF\_SN\_AGN} reproduces the observed low baryon fractions of galaxy groups (see Paper I).

\subsection{Excess entropy in runs without SN and AGN feedback}

Before discussing the Lagrangian entropy distributions and histories, we present Fig.\ 3, which shows the gas, stellar, and total baryon mass fractions within $r_{500}$ as a function of mass, $M_{500}$, of the simulated groups in all of the runs we consider in the present study.  This figure is useful for the interpretation of the results presented below and will be referred to frequently.

\subsubsection{Present-day entropy distributions}

In Fig.\ 4 we plot the $z=0$ sorted gas entropy as a function of the included gas mass fraction for the {\it NoCool}, {\it PrimCool\_SF} and {\it ZCool\_SF} runs.  The sorted entropy is calculated for each group by ordering the hot ($T > 10^5$ K) gas particles within $r_{500}$ by their entropies and calculating the total mass of gas, $M_{\rm gas}^{'}(<S)$, with entropy lower than $S$.  The motivation for examining $S(M_{\rm gas}^{'})$, rather than $S(r)$, is that cooling modifies $S(M_{\rm gas}^{'})$ in a very simple way: it selectively removes the gas with entropy lower than the cooling threshold (see VB; see also below).  The resulting radial entropy distribution, by contrast, depends not only on the resulting entropy, but also on the depth of the gravitational potential well, since the gas will tend to re-establish hydrostatic and convective equilibrium.  For this reason, it is conceptually much easier to deduce the effects of cooling on the \igm\ using the sorted gas entropy as a function of the included gas mass fraction.  The solid curves represent the median relations for the three runs for groups with $13.25 \le \log_{10}(M_{500}/M_\odot) \le 14.25$.

\begin{figure}
\includegraphics[width=\columnwidth]{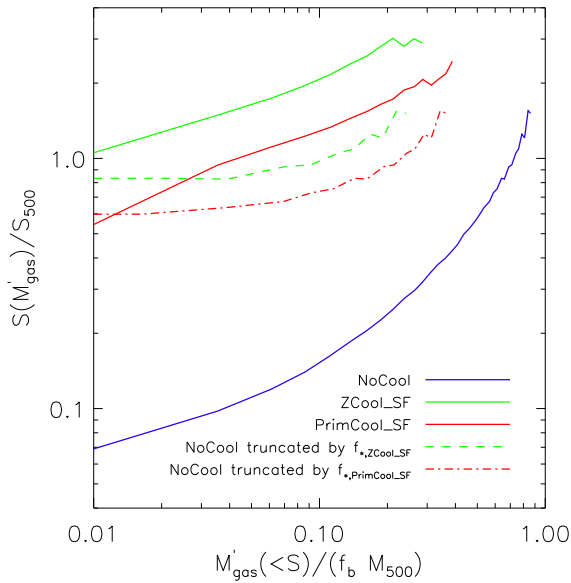}
\caption{The sorted gas entropy as a function of included gas mass fraction for the {\it NoCool}, {\it ZCool\_SF} and {\it PrimCool\_SF} runs (solid blue, green, and red curves, respectively). This is calculated for each group by ordering the gas particles within $r_{500}$ by their entropies and calculating the total mass of gas with entropy lower than $S$, $M_{\rm gas}^{'}(<S)$.  The solid curves represent the median relations for the 3 runs for groups with $13.25 \le \log_{10}(M_{500}/M_\odot) \le 14.25$.  The dashed green and red curves represent the sorted entropy as a function of enclosed gas mass fraction for the {\it NoCool} run when all gas with entropy lower than that which encloses $f_{\rm gas} = f_{\rm *,ZCool\_SF}$ and $f_{\rm *,PrimCool\_SF}$, respectively, is removed (``truncated'').  This operation yields the expected entropy distribution after cooling has selectively removed the lowest entropy/shortest cooling time gas.  Overall, the truncated entropy distributions are similar to those obtained directly from {\it ZCool\_SF} and {\it PrimCool\_SF}, indicating that the excess entropy in these runs is indeed largely the result of selective removal of low entropy gas, as proposed by VB.  See text for further discussion.
}
\label{fig:entropy_mgas_nocool_nosn}
\end{figure}

Note that for {\it NoCool} the gas mass fraction within $r_{500}$ is approximately $0.9$ times the universal baryon fraction, $f_b$ (the blue curve ends at $M_{\rm gas}^{'}(<S) \approx 0.9 f_b$ M$_{500}$).  This is in accordance with the results of previous non-radiative SPH simulations which calculated the baryon fraction within a similar fixed aperture (e.g., Eke et al.\ 1998; Frenk et al.\ 1999; Kravtsov et al.\ 2005; Crain et al.\ 2007).  The gas mass fractions for {\it ZCool\_SF} and {\it PrimCool\_SF}, however, are much lower - approximately $0.3$ and $0.4$ times the universal baryon fraction, respectively (see top panel of Fig.\ 3).  In these runs most of the baryons have been able to cool significantly and form stars (middle panel of Fig.\ 3).  However, the {\it total} baryon fraction (stars + gas) is very similar to that of {\it NoCool}, as can be seen from the bottom panel of Fig.\ 3.  Kravtsov et al.\ (2005) find very similar trends for their simulations that include radiative cooling and star formation (and inefficient feedback).

A comparison of the solid curves reveals excess entropy at fixed gas mass fraction in the runs with cooling relative to the baseline {\it NoCool} run.  This plot is similar to the trend seen in Fig.\ 2 because the buoyancy of the gas results in a monotonic mapping between $r$ and $M_{\rm gas}^{'}(<S)$.  According to the model proposed by VB, cooling selectively removes the lowest entropy gas from the hot phase, which results in a shift in the $S(M_{\rm gas}^{'})$ curves to the left in Fig.\ 4.  To see whether this idea works in detail, we truncate the {\it NoCool} $S(M_{\rm gas}^{'})$ distribution (i.e., shift to the left) using the calculated stellar fractions from the simulated groups in {\it ZCool\_SF} and {\it PrimCool\_SF} (dashed and dot-dashed curves, respectively).  This is done by removing all gas with entropies lower than the entropy that encloses $f_{\rm gas} = f_{\rm *}$.  The remaining gas mass fraction is then just the original gas mass fraction minus the stellar mass fraction.  This procedure yields the expected $S(M_{\rm gas})$ distribution after cooling has removed the lowest entropy gas from the hot phase\footnote{More precisely, this represents an {\it upper limit} to the entropy distribution after cooling has selectively removed the lowest entropy gas, as it has been implicitly assumed that any gas that has not cooled out of the hot phase flows adiabatically inward to replace the gas that has cooled out.  In reality, however, some of the remaining hot gas would have its entropy reduced by cooling as its density increases.}.

Overall, the entropy distributions derived by truncating the {\it NoCool} distribution using the stellar fractions from {\it ZCool\_SF} and {\it PrimCool\_SF}  are similar to the actual entropy distributions of the hot gas obtained from these runs.  This indicates that truncation due to cooling is the likely explanation for the bulk of the excess entropy in these runs.  But note that a shortfall of a factor of approximately 1.5 at intermediate gas masses is present, implying that the VB truncation mechanism is not a perfect description of what happens in the simulations.  A plausible explanation for this shortfall is that the remaining hot gas in {\it ZCool\_SF} and {\it PrimCool\_SF} is (gravitationally) shocked to a higher degree than the same gas in the {\it NoCool} run because it is able to fall further into the gravitational potential of the group, and therefore shock at a higher Mach number, due to the reduced baryon fraction (and thermal pressure) of the group (e.g., McCarthy et al.\ 2007).  Indeed, as we discuss below, the gas that constitutes the \igm\ in {\it ZCool\_SF} and {\it PrimCool\_SF} is largely located at $r > r_{500}$ in {\it NoCool}, implying thermal pressure has held the gas up in the latter.

\subsubsection{Lagrangian thermal histories}

To test these ideas, we calculate for each group the median entropy, density, and temperature as a function of redshift for all the gas that ends up hot ($T > 10^5$ K) and within $r_{500}$ at the present day (i.e, that which constitutes the IGrM).  In Fig.\ 5 we show the median histories for galaxy groups in the mass range $13.25 \le \log_{10}(M_{500}/M_\odot) \le 14.25$.  The solid blue and red curves represent {\it NoCool} and {\it ZCool\_SF}, respectively.

Focusing first on the entropy histories (top panel), we see that prior to $z\approx2$ the two runs are quite similar; they show a gradual rise in the gas entropy due to both gravitational shock heating (i.e., particles falling into potential wells, which are becoming deeper with time) and UV photoheating.  The sharp rise in entropy at $z=9$ corresponds to reionisation, which is assumed to occur everywhere in the simulation volume simultaneously.  At late times, the entropy histories diverge, such that by the present day ($z=0$) there is approximately a factor of $4$ difference in the medians, consistent with the $z=0$ entropy profiles plotted in Fig.\ 2.  
However, it is important to note that by selecting the hot gas within $r_{500}$ at $z=0$ for the two simulations, we are not selecting identical sets of particles.  The gas that ends up constituting the \igm\ in one simulation may differ from that which ends up constituting the \igm\ in another simulation.  This implies that we are comparing the thermal histories of different gas `parcels'.  However, by using the unique IDs of the gas particles, we can select identical sets of particles in the simulations and compare their Lagrangian thermal histories (note by using massless tracer particles, similar analyses can also be applied to mesh-based simulations, e.g., Vazza 2010).

\begin{figure}
\includegraphics[width=\columnwidth]{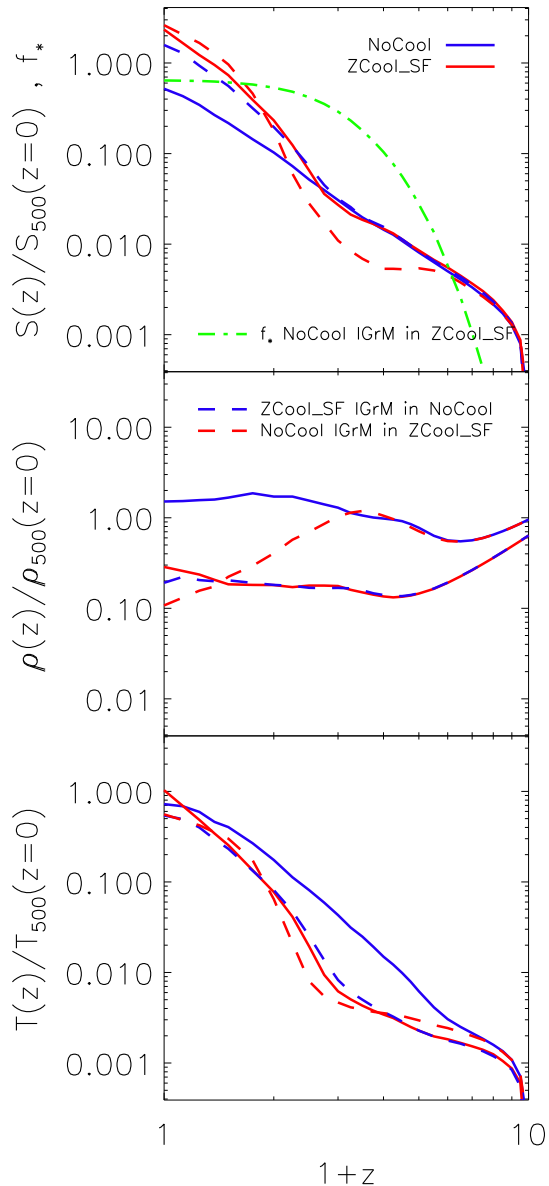}
\caption{{\it Top:}
The solid blue and red curves represent for {\it NoCool} and {\it ZCool\_SF}, respectively, the median entropy history of gas that ends up hot and within $r_{500}$ at the present day for groups in the mass range $13.25 \le \log_{10}(M_{500}/M_\odot) \le 14.25$.  The dashed blue curve represents the median entropy history of gas in {\it NoCool} that ends up hot and within $r_{500}$ in {\it ZCool\_SF}.  The dashed red curve represents the median entropy history of gas in the {\it ZCool\_SF} run that ends up hot and within $r_{500}$ in the {\it NoCool} run.  The dot dashed green curve represents the fraction of group gas particles within $r_{500}$ in {\it NoCool} that have been converted into star particles in {\it ZCool\_SF}.  {\it Middle:} The analogous plot for the gas density.  The gas density has been normalised by the `virial gas density' $\rho_{500}(z=0) \equiv 500 f_b \rho_{\rm crit}(z=0)$.  {\it Bottom:}  The analogous plot for the gas temperature.  The temperature has been normalised by the `virial temperature' $T_{500}(z=0) \equiv G M_{500} \mu m_H / (2 r_{500} k_B)$.  When identical sets of particles are selected in the two runs (compare the solid red and dashed blue curves) the histories are similar.  This implies that the excess entropy in the {\it ZCool\_SF} run is mostly due to the selective removal of the lowest entropy gas via cooling and star formation.}
\label{fig:thermal_history_nocool_nosn}
\end{figure}

The dashed blue curve in the top panel of Fig.\ 5 shows the median entropy history of gas in the {\it NoCool} (non-radiative) run that ends up constituting the \igm\ in {\it ZCool\_SF} (i.e., is hot and within $r_{500}$ at $z=0$ in {\it ZCool\_SF}).  (But note that there is no requirement that this gas ends up as part of the \igm\ in the {\it NoCool} run itself.)
By comparing this curve to the solid red curve, we are making a comparison of identical sets of particles in two runs.  As can clearly be seen, the histories are remarkably similar, with the median entropies differing only by a factor of $1.5$ or so at the present day.  This agrees with the difference between the solid and dashed green curves seen in Fig.\ 4.
Since we selected and followed back in time the same set of particles, which are those particles that end up hot and in groups at $z=0$ in {\it ZCool\_SF}, the similarity of the histories firmly establishes that the VB truncation mechanism is responsible for the {\it bulk} of the excess entropy in {\it ZCool\_SF}.  We can make this statement as we have selected only those particles that {\it survived} truncation.  The similarity of the histories implies that the reason for the large difference in the $z=0$ (non-truncated) entropy distributions is a result of cooling selectively-removing the lowest entropy gas.  As also predicted by VB, the remaining hot gas flows inward to replace the truncated low entropy gas.  The median radius of the hot gas within $r_{500}$ of {\it ZCool\_SF} is approximately $0.7 r_{500}$, whereas the same set of gas particles have a median radius of approximately $1.3 r_{500}$ in {\it NoCool}.  Hot gas in {\it NoCool} is unable to cool and its pressure prevents gas outside of $r_{500}$ from falling inward.

We can also perform the inverse test, which is to select from the {\it NoCool} run those particles that end up hot and within $r_{500}$ of galaxy groups and show their history in the {\it ZCool\_SF} run.  This is represented by the dashed red curve in the top panel of Fig.\ 5.  Of course, a large fraction of the gas that ends up hot and within $r_{500}$ in the {\it NoCool} run will be able to cool and form stars in the {\it ZCool\_SF} run.  The dashed red curve therefore represents the median entropy of the gas which does not get converted into stars (therefore the number of gas particles we use to calculate the median entropy changes with time for this curve).  The dot-dashed green curve represents the fraction of gas particles (that end up hot and within $r_{500}$ in the {\it NoCool} run) that are converted into star particles in the {\it ZCool\_SF} run.

At early times, $z \ga 5$, cooling is unimportant for the gas in the {\it ZCool\_SF} run that ends up constituting the \igm\ in the {\it NoCool} run and, as a result, the entropy history of this population of particles in the {\it ZCool\_SF} run is similar to that of the same gas in the {\it NoCool} run.  At $z \la 5$, cooling becomes rapidly more important and we see a departure between the populations in the two runs, with the gas in the {\it ZCool\_SF} run having stalled its entropy production.  Not coincidentally, this departure corresponds closely to the time when the gas begins to be rapidly converted into stars.  Eventually, by $z \sim 1$ all of the low entropy gas is converted into stars (see dot-dashed green curve).  The entropy history of the remaining hot gas then becomes very similar to that represented by the solid red curve, since these are now almost identical sets of particles.

All of the above information was gleaned from an analysis of the entropy history alone, but it is worthwhile to briefly consider the density and temperature histories as well.  In the middle and bottom panels of Fig.\ 5 we see that gas that is destined to constitute the \igm\ in the {\it ZCool\_SF} run (solid red) is typically cooler (at least until recently) and less dense than that which will constitute the \igm\ in the {\it NoCool} run (solid blue).  Given the similarity of the entropy histories of the two runs until $z \approx 2$, this suggests the \igm\ in the {\it NoCool} run was, on average, compressed (adiabatically) more strongly at early times, presumably because it was typically located deeper within the potential wells of small haloes.  Gas that is destined to constitute the \igm\ in the {\it ZCool\_SF} run, by contrast, is that which preferentially avoided significant cooling due to its lower density, presumably because it was located more on the periphery of overdensities at early times.  This is confirmed by examining the dashed blue curve.

It is also worth pointing out that, had we focused on the density and/or temperature alone without consideration for the entropy, we might have reasonably (but wrongly) deduced that cooling was directly important at very early times ($z \ga 5$), since the densities and temperatures of the gas in the two runs differ at that time.  However, the similarity of the entropies of the gas demonstrates that this is due purely to differences in the (adiabatic) compression of the gas at early times.

We have also examined the thermal history of the other run without feedback ({\it PrimCool\_SF}) and find that it is very similar to that of {\it ZCool\_SF} and confirms the above conclusions.

\subsection{Excess entropy in runs with SN feedback}

\begin{figure}
\includegraphics[width=\columnwidth]{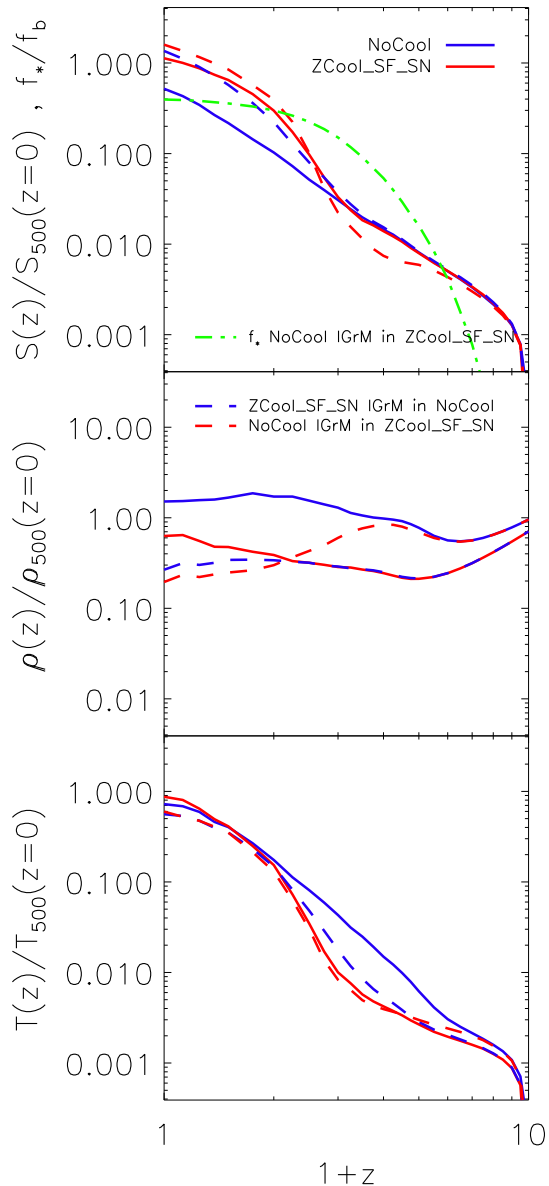}
\caption{Same as Fig.\ 5 but for {\it ZCool\_SF\_SN}.  The similarity of the solid red and dashed blue curves indicates that SN feedback has not significantly raised the entropy of the gas that constitutes the present-day \igm\ in {\it ZCool\_SF\_SN}.  The similarity of the dashed red and solid red curves indicates that SN feedback has not strongly heated any gas that may have been ejected either. }
\label{fig:thermal_history_nocool_ref}
\end{figure}

In Section 3.2 we demonstrated that in runs without feedback the excess entropy at intermediate radii is largely a consequence of the selective removal of low entropy gas due to cooling and star formation.  Runs that include SN feedback also show evidence for excess entropy relative to the self-similar model (see Fig.\ 2), but it is not immediately obvious in this case whether it is cooling/star formation or SN heating (or both) that is responsible for the extra entropy.  A comparison of the thermal history of the groups in runs with SN feedback with that of {\it NoCool} can resolve this issue.

In Fig.\ 6 we show the median entropy, density, and temperature histories for galaxy groups in the mass range $13.25 \le \log_{10}(M_{500}/M_\odot) \le 14.25$.  The solid blue and red curves represent {\it NoCool} and {\it ZCool\_SF\_SN}, respectively.  Focusing first on the entropy (top panel), at the present day ($z=0$) there is approximately a factor of $2$ difference between the medians in the two runs.  This is smaller than the difference we found between {\it NoCool} and {\it ZCool\_SF} (Fig.\ 5).  This is already a clue that the addition of SN feedback does not significantly raise the entropy of the IGrM.

Selecting from {\it NoCool} only those particles which end up being hot and within $r_{500}$ in {\it ZCool\_SF\_SN} (dashed blue curve), we see a remarkably similar thermal history to this same particle population in {\it ZCool\_SF\_SN} (solid red curve).  This immediately tells us that SN feedback has not significantly raised the entropy of gas that ends up hot and within groups in {\it ZCool\_SF\_SN}.  It is conceivable, though, that some of the gas {\it could} have been heated significantly by SN feedback and, as a result, did not end up within $r_{500}$ in {\it ZCool\_SF\_SN}.  However, the similarity of the dashed red curve (i.e., those gas particles in {\it ZCool\_SF\_SN} that end up hot and within groups in {\it NoCool}) and the solid red curve (gas that ends up hot and in groups in the {\it ZCool\_SF\_SN} run) demonstrates that SN feedback did not significantly raise the entropy of any gas ever associated with the groups in {\it ZCool\_SF\_SN}.

As in the case of the runs without SN feedback, we thus find that the primary cause of excess entropy in {\it ZCool\_SF\_SN} is the selective removal of low entropy gas.  As already noted, the amount of excess entropy in this run is smaller than for runs without feedback.  In addition, the stellar fraction in this run, $f_*(r_{500}) \approx 0.4 f_b$, is smaller than that for the runs without SN feedback ($\approx 0.6 f_b$ for {\it ZCool\_SF}; see Fig.\ 3).  The role that SN feedback plays, therefore, is to prevent some of the low entropy gas from turning into cold gas/stars which implies that the entropy distribution is not truncated to the same degree as in the runs with no effective feedback.  Consequently, the entropy of the \igm\ is not raised by the same amount.  And feedback from SN-driven winds does not significantly heat the IGrM.  It is worth noting here that since the SN feedback is implemented in kinetic form in the \owls\ simulations (see Dalla Vecchia \& Schaye 2008), it does not suffer from the standard problem associated with most thermal implementations of SN feedback; i.e., that the gas radiates away its gained energy almost immediately.

The density and temperature histories yield a similar story to what we found by analysing the runs with no feedback; i.e., the gas that constitutes the \igm\ at the present-day in the {\it ZCool\_SF\_SN} run is that which escaped significant cooling because it was of lower density and temperature at early times.  Again, this is presumably because it was located in the outskirts of haloes at early times.

The thermal history of the other run with SN feedback we consider (i.e., {\it PrimCool\_SF\_SN}) is very similar to that of {\it ZCool\_SF\_SN}, in the sense that SN feedback does not noticeably raise the entropy of the IGrM.  Instead, it acts primarily to prevent some of the lowest entropy gas from turning into stars/cold gas.  However, the lack of metal-line cooling in this run greatly reduces the fraction of baryons converted into stars, resulting in a smaller truncation of the entropy distribution and entropy profiles that clearly are at odds with observations (see Fig.\ 2).

\subsection{Excess entropy in a run with AGN feedback}

\begin{figure}
\includegraphics[width=\columnwidth]{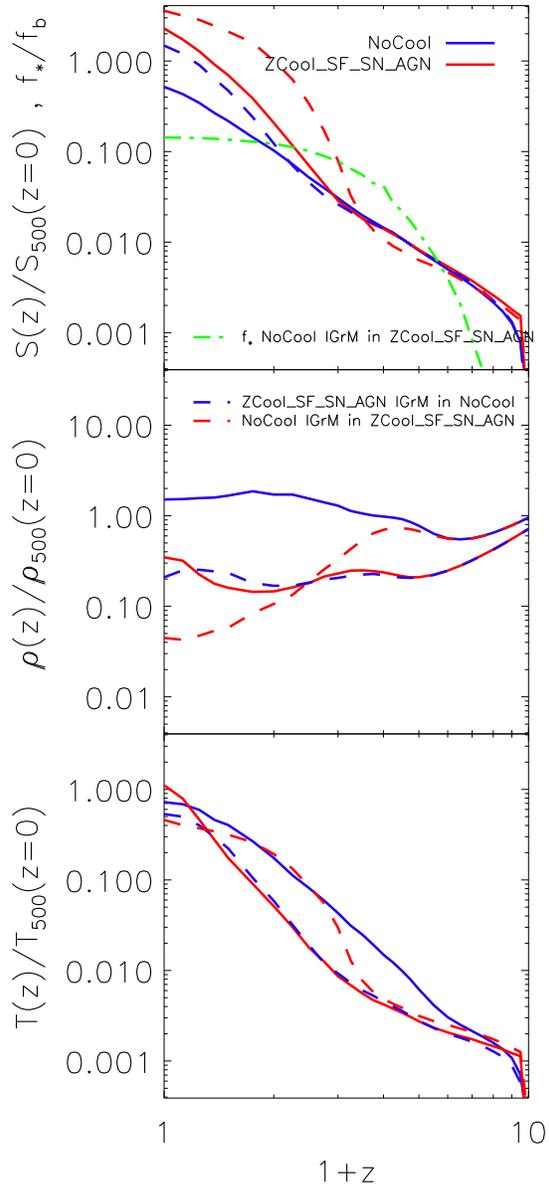}
\caption{Same as Fig.\ 5 but for {\it ZCool\_SF\_SN\_AGN}.  The similarity of the solid red and dashed blue curves indicates that {\it feedback from BHs has not strongly heated the gas that ends up constituting the present-day IGrM.}  However, the deviation of the dashed red and solid red curves at $z \approx 2$ implies that AGN feedback has significantly heated gas which did not end up in groups in {\it ZCool\_SF\_SN\_AGN} but did in {\it NoCool} (i.e., ejected gas).}
\label{fig:thermal_history_nocool_agn}
\end{figure}

The \owls\ runs we have investigated so far, particularly those which include the important process of metal-line cooling, produce galaxy group stellar mass fractions that are much larger than observed.  In particular, current estimates suggest that $f_*(r_{500})$ is between 0.1-0.2 $f_b$ for nearby groups (Lin \& Mohr 2004; Balogh et al.\ 2008; McGee \& Balogh 2010), depending on what fraction of the total light is assumed to be in a diffuse component (see, e.g., Gonzalez et al.\ 2007; McGee \& Balogh 2010).  This implies that {\it ZCool\_SF\_SN} and {\it ZCool\_SF} have too high stellar mass fractions by factors of $\approx$ 3 and 4, respectively (see Fig.\ 3).  This is the so-called `cooling crisis' of cosmological hydrodynamic simulations that invoke inefficient feedback (e.g., Balogh et al.\ 2001).  

Based on the above results, any simulation that has too high a stellar mass fraction will overestimate the degree to which the entropy of the \igm\ will be raised by cooling/star formation, because it overestimates the degree to which the self-similar entropy distribution has been truncated.  It is interesting to note that of the runs we have examined so far, the only ones with excess entropy comparable to what is observed ({\it ZCool\_SF}, {\it PrimCool\_SF}, {\it ZCool\_SF\_SN}) significantly violate the observed stellar mass fraction of groups.  It is also interesting that while the model that includes SN feedback but calculates cooling rates assuming primordial abundances ({\it PrimCool\_SF\_SN}), predicts a stellar mass fraction only slightly larger than observed, the predicted entropy is much lower than observed\footnote{Of course {\it PrimCool\_SF\_SN} ignores the important effects of metal-line cooling, so it does not represent a physically reasonable model.  We note, however, than many cosmological simulations found in the literature adopt primordial abundances when calculating cooling rates.}.  It therefore does not appear possible to simultaneously match the observed thermal properties of the \igm\ and the observed stellar mass fractions of groups in the context of a model where the entropy is ``raised'' (by truncation of low entropy gas) via cooling/star formation alone.  This likely indicates that some form of highly efficient feedback (in terms of raising the entropy) is required in order to explain the excess entropy and stellar mass fractions of galaxy groups simultaneously.  

\begin{figure}
\includegraphics[width=\columnwidth]{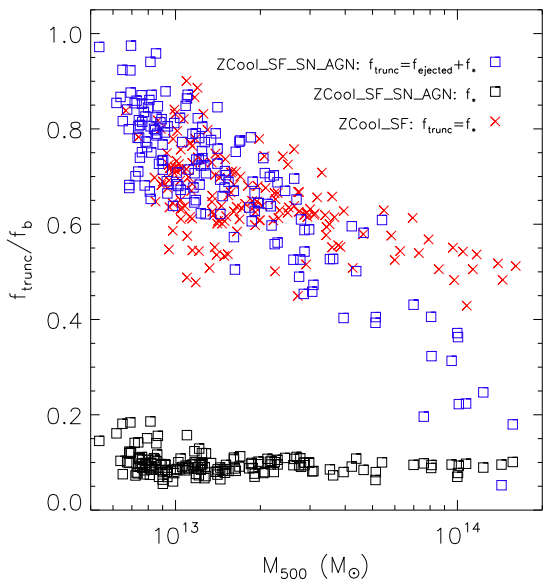}
\caption{The fraction of gas removed (truncated) from the hot \igm\ due to star formation in {\it ZCool\_SF} or star formation plus ejection in {\it ZCool\_SF\_SN\_AGN} as a function of group mass.  Also shown is the fraction removed from the hot \igm\ due to star formation alone in {\it ZCool\_SF\_SN\_AGN}.
The reason the two runs have similar present-day entropy at intermediate radii is that both target the same low-entropy gas for truncation. }
\label{fig:entropy_agn_nosn_mass}
\end{figure}

We demonstrated in Paper I that model {\it ZCool\_SF\_SN\_AGN}, which includes AGN feedback, yields the correct stellar mass fraction and entropy distribution for the \igm\ as a function of halo mass, in addition to matching a wide range of additional properties.  What is the physical process that sets the entropy distribution of the \igm\ in this model? Four possible scenarios are as follows:  

\begin{itemize}
\item{BH feedback raises the entropy of the gas which, in turn, increases its cooling time and reduces its star formation efficiency.  Since the peak of the star formation (and presumably BH formation) history of the universe occurs at $z \approx 2-3$, and present-day groups contain most of the galaxies, this scenario would imply that the excess entropy was generated at high redshift, well before the formation of the IGrM.  This is another way of saying the \igm\ was `preheated' (e.g., Kaiser 1991; Evrard \& Henry 1991).}

\item{A second possibility is that the entropy was raised much more recently through direct heating by BHs\footnote{Although McCarthy et al.\ (2008) argued that such internal heating is energetically prohibitive due to the high density of the gas after the system has collapsed.}.}  

\item{A third possibility is that BH feedback targets only the lowest entropy/shortest cooling time gas for feedback, ejecting it from the low-mass progenitors of present-day groups, while leaving the long cooling time gas unaltered (i.e., the long cooling time gas ends up forming the present-day IGrM).  Here we do not distinguish whether the low entropy gas was first heated and then buoyantly floated out of the system or if it was launched balistically (i.e., given a large velocity kick) from the system.  But note the entropy gain/velocity kick must be sufficiently large so that the gas is not re-accreted as the group halo grows.  Some tuning of this mechanism is needed, as the gas {\it is} recaptured by massive clusters (i.e., their baryon fractions are near the universal value).}

\item{Low entropy gas is ejected from present-day groups, as proposed recently by B08.  To be more accurate, B08 actually allowed for ejection from the low mass-progenitors of groups as well.  However, this gas is re-accreted by the group at late times if the energy imparted to the gas is less than the binding energy of the present-day group.  Therefore, the energetics required to eject the gas are set by the depth of the potential well of the group at $z=0$, which is why we refer to B08 as a `present-day ejection' model.} 
\end{itemize}

The last two scenarios differ from the first two in that the excess entropy in groups is not achieved by heating of the \igm\ but instead by the removal of the lowest entropy gas, as proposed by VB, modulo that the truncation results from the combination of cooling/star formation and ejection.

An examination of the thermal history of {\it ZCool\_SF\_SN\_AGN} can distinguish between these four proposed scenarios (i.e., early heating, late heating, early ejection, and late ejection, respectively).  In Fig.\ 7 we show the median entropy, density, and temperature histories for galaxy groups in the mass range $13.25 \le \log_{10}(M_{500}/M_\odot) \le 14.25$.  The solid blue and red curves represent {\it NoCool} and {\it ZCool\_SF\_SN\_AGN}, respectively.  Focusing first on the entropy (top panel), at the present day ($z=0$) there is approximately a factor of $4$ difference between the medians in the two runs, which is similar to the difference between {\it ZCool\_SF} and {\it NoCool} (Fig.\ 5).  We discuss below whether or not this is a coincidence.

Selecting in the {\it NoCool} run only those particles that end up hot and within $r_{500}$ in {\it ZCool\_SF\_SN\_AGN} (dashed blue curve), we see a similar thermal history as this same particle population has in {\it ZCool\_SF\_SN\_AGN} (solid red curve).  This is somewhat surprising as it indicates that {\it feedback from supermassive BHs has not substantially raised the entropy of gas that ends up constituting the IGrM} in the run with AGN feedback\footnote{Note a slight offset {\it is} present between the solid red and dashed blue curves. However, as we discuss in Section 4.2, this is most likely not a consequence of AGN feedback but, rather, amplified gravitational shock heating due to the reduced thermal pressure of the \igm\ in the groups in the run with AGN.}.  This immediately rules out the first two scenarios we outlined above, which are based on the idea that the excess entropy in groups is a result of heating of the gas that ends up in the IGrM.

Of course, some amount of gas could have been heated significantly by AGN feedback and, as a result, ended up outside of $r_{500}$ in {\it ZCool\_SF\_SN\_AGN}.  Indeed, the reduced baryon fractions of groups in the {\it ZCool\_SF\_SN\_AGN} run relative to the universal value (see Fig.\ 3) already indicates that this is the case.  This is reflected in the entropy history as a departure of the dashed red curve (i.e., those particles in {\it ZCool\_SF\_SN\_AGN} that end up hot and within groups in {\it NoCool}) from the solid red curve (gas that ends up hot and in groups in {\it ZCool\_SF\_SN\_AGN}) at $z \approx 2$.  BH feedback has therefore raised the entropy of some of the gas that would have ended up in groups if not for this boost.  This is contrary to what we found for the runs with only SN feedback (see Fig.\ 6).

Thus, the way in which the entropy of the \igm\ is raised by feedback from BHs is not by heating of gas destined to end up in groups, but instead by ejecting low entropy gas at early times.  We noted above that the degree to which the entropy is raised in {\it ZCool\_SF\_SN\_AGN} is similar to that in the run that includes metal-line cooling and star formation but no significant sources of feedback ({\it ZCool\_SF}).  At first sight this would appear to be a coincidence.  However, this is exactly what one expects if the excess entropy is due primarily to the removal of just the lowest entropy gas (which happens via star formation in {\it ZCool\_SF} and star formation plus ejection in {\it ZCool\_SF\_SN\_AGN}).  We therefore expect that the sum of the stellar mass and ejected gas mass fractions for {\it ZCool\_SF\_SN\_AGN} to be similar to the stellar mass fraction in {\it ZCool\_SF}, which is something that can be explicitly checked in the simulations.

In Fig.\ 8 we show the fraction of gas that has been truncated as a function of M$_{500}$ for {\it ZCool\_SF\_SN\_AGN} and {\it ZCool\_SF}.  The truncation fraction in {\it ZCool\_SF} is simply the stellar mass fraction, whereas for {\it ZCool\_SF\_SN\_AGN} it is the summation of the stellar mass fraction and the ejected gas mass fraction.  We compute the ejected gas mass fraction for each group in the {\it ZCool\_SF\_SN\_AGN} as the difference in the baryon fraction of the same group between {\it ZCool\_SF} and {\it ZCool\_SF\_SN\_AGN}.

We see that over a large range in halo mass (M$_{500} \la 5\times10^{13} M_\odot$) the degree of truncation in the two simulations is indeed quite similar, providing strong evidence that it is indeed the lowest entropy gas that has been targeted by BH feedback.  Any low entropy gas that escapes BH feedback ends up forming stars, but clearly it is ejection, rather than star formation, that dominates the truncation fraction, as can be quickly deduced by comparison to the stellar mass fractions of the {\it ZCool\_SF\_SN\_AGN} run.

Interestingly, at high halo masses the truncation fractions of the two runs begin to depart, with $f_{\rm trunc}$ approaching $f_{\rm *}$ for {\it ZCool\_SF\_SN\_AGN}.  This is not unexpected, as it becomes increasingly more difficult for the BH feedback to eject gas beyond $r_{500}$, due to the increased binding energies of more massive systems (see also Fig.\ 3, which shows the total baryon fraction is approaching the universal value for these very massive systems).

\begin{figure}
\includegraphics[width=\columnwidth]{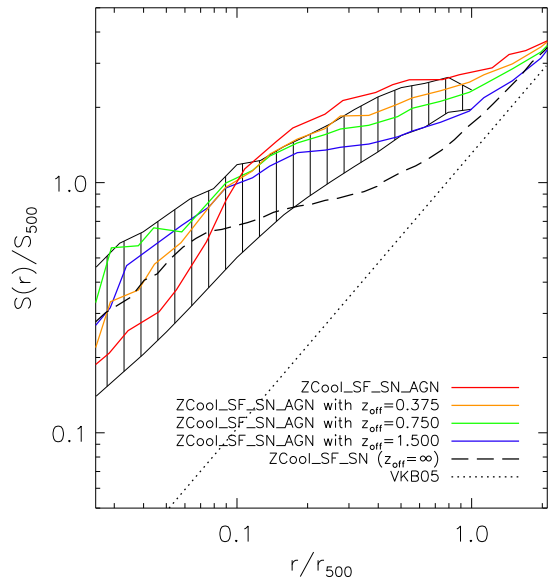}
\caption{The timing of entropy injection due to AGN feedback in the {\it ZCool\_SF\_SN\_AGN} run.  Shown are the median 3D mass-weighted entropy profiles at $z=0$ for the AGN feedback model when the AGN have been switched off in the simulation at redshift $z=z_{\rm off}$ (see text).  The median is calculated for groups in the mass range $13.25 \le \log_{10}(M_{500}/M_\odot) \le 14.25$.  For reference, the hatched region represents the observational data of Sun et al.\ (2009). 
 Feedback from AGN-driven outflows has significantly heated much of the gas that had the lowest entropy at high redshift, ejecting it from groups primarily at $z \ga 1.5$.}
\end{figure}

\section{The Physical nature of AGN feedback}

We demonstrated in Section 3.4 that the way in which feedback from AGN raises the entropy of the \igm\ is through the ejection of low entropy gas, rather than heating of the gas that constitutes the present-day IGrM.  This interesting result warrants further investigation.  In particular, we would like to know how, when, and to what degree the ejected gas is heated by feedback from BHs.  We address these questions below.  We also briefly discuss the potential sensitivity of our conclusions to the adopted parameters of the AGN feedback model.

\subsection{When was the gas ejected/heated?}

In Fig.\ 9 we show the median radial entropy distribution of the gas for runs using the AGN model where AGN feedback has been explicitly shut off in the simulation at redshift $z=z_{\rm off}$ (i.e., we have re-run the AGN model a number of times, each time explicitly switching off AGN feedback at a different redshift).  Switching off the AGN in the simulation provides a simple tool for deducing the importance of AGN feedback subsequent to $z_{\rm off}$.  The simulation with metal-dependent cooling, star formation, and feedback from SNe but not from AGN ({\it ZCool\_SF\_SN}) would correspond to the case of $z_{\rm off}=\infty$ in this terminology.

\begin{figure}
\includegraphics[width=\columnwidth]{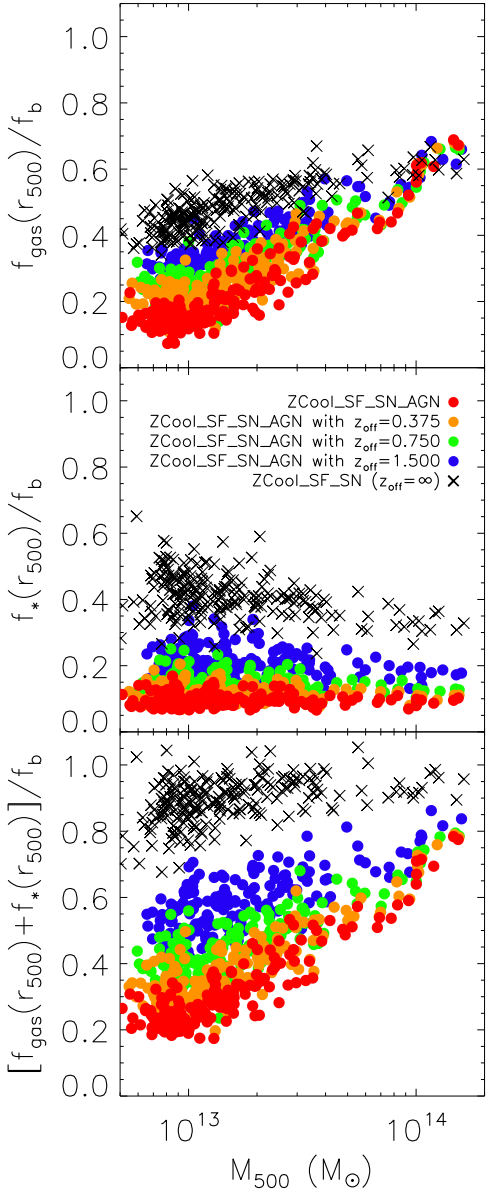}
\caption{Gas mass (top), stellar mass (middle), and total baryon (bottom) mass fractions within $r_{500}$ as a function of group mass, M$_{500}$, for the AGN feedback model when the AGN have been switched off in the simulation at redshift $z=z_{\rm off}$.  The mass fractions have been normalised to the universal baryon fraction, $f_b = \Omega_b/\Omega_m$.  Feedback from AGN-driven outflows has significantly heated much of the lowest entropy gas, ejecting it from groups primarily at $z \ga 1.5$.}
\end{figure}

Comparing the dashed black curve ($z_{\rm off}=\infty$) and solid blue ($z_{\rm off}=1.5$) and red (AGN never shut off) curves at intermediate radii, we see that the bulk of the excess entropy relative to the self-similar profile is produced at $z \ga 1.5$, since the solid blue curve is about half way between the dashed black and solid red curves.  Note that this is likely a lower limit for the redshift where AGN feedback raises the entropy of the IGrM, since excess entropy can be produced even after the AGN is switched off via truncation due to radiative cooling (as in Section 3.3). 

We show the effect of switching off the AGN on the gas, stellar, and baryon mass fractions within $r_{500}$ at $z=0$ in Fig.\ 10.  Consistent with the entropy profiles shown in Fig.\ 9, we see here that the bulk of the ejection of gas (and the prevention of star formation) through AGN feedback occurred at $z \ga 1.5$, as the blue filled circles (representing the case $z_{\rm off} = 1.5$) are closer to the solid filled circles ({\it ZCool\_SF\_SN\_AGN}) than to the black crosses ($z_{\rm off}=\infty$) in all three panels shown in Fig.\ 10.  The AGN model solves the overcooling problem common to many previous cosmological simulations {\it because} the feedback is efficient at high redshift, which is when the bulk of the star formation occurs (e.g., Nagai \& Kravtsov 2004).

A more rigorous, though perhaps less intuitive, determination of the epoch during which the gas was heated/ejected can be deduced from the median entropy history of ejected gas in {\it ZCool\_SF\_SN\_AGN}, which is shown in Fig.\ 11.  Also shown (solid black) is the median entropy history of the gas that ends up constituting the \igm\ in this run.

\begin{figure}
\includegraphics[width=\columnwidth]{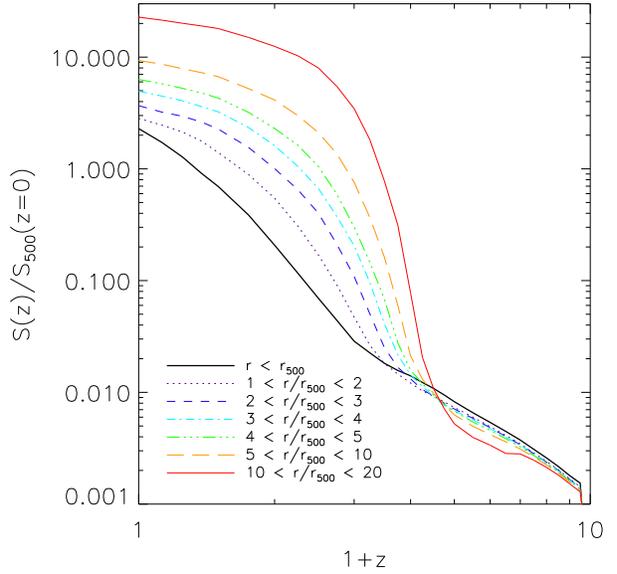}
\caption{The median entropy history of ejected gas in {\it ZCool\_SF\_SN\_AGN}. (An ejected gas particle is one that is located within $r_{500}$ in {\it NoCool} but is outside this radius in {\it ZCool\_SF\_SN\_AGN}.)
 Feedback from AGN-driven outflows has significantly heated much of the gas that had the lowest entropy at high redshift, ejecting it from groups primarily at $z \ga 1.5$.}
\end{figure}

At early times ($z \ga 4$) the entropy history of ejected gas is similar to that of gas which is not ejected.  This indicates that feedback from BHs is unimportant prior to $z \approx 4$.  However, over a relatively small range of redshift ($2 \la z \la 4$), we see a sharp departure of the ejected gas from the gas destined to constitute the IGrM, indicating that AGN feedback is very important during this time.  This behaviour is what we naively expect given the mass accretion rate history of the model (see Fig.\ 7 of Booth \& Schaye 2009, which shows
that the BHs gain most of their mass between $z=4$ and $z=2$).   
Observationally, this is the epoch when the quasar luminosity function 
peaks, and the energy injection rate is proportional to the luminosity.
It is important to note that most of the BH growth occurs when the accretion rate is close to Eddington-limited, which is when the BH would be considered to be in `quasar mode', rather than in `radio mode'.  That our simulations indicate that the entropy of the present-day \igm\ is largely shaped by AGN `quasar mode' feedback differs from previous studies (e.g., B08; Giodini et al.\ 2010) and is one of our major results.  Note that the time period over which the heating/ejection occurs also corresponds to the peak of the star formation rate density of the universe for this run (see Booth \& Schaye 2009), indicating that there is a strong correlation between BH growth and star formation in the simulation, as is also observed (e.g., Boyle \& Terlevich 1998; Kauffmann et al.\ 2003; H{\"a}ring \& Rix 2004).  Indeed, this is presumably the reason why the simulation reproduces the $z=0$ black hole mass - stellar mass relation.

The ordering of the curves in the right panel of Fig.\ 11 is also very interesting.  It indicates that the lowest entropy gas is the first to be ejected and ultimately ends up gaining the most entropy.  This is consistent with the expectation that the gas that is ejected early on is able to travel further away from the group by $z=0$ and will therefore have acquired more entropy through weak shocking as it interacts with the intergalactic medium.

\subsection{To what degree is the ejected gas heated?}

To more clearly demonstrate the effect of BH feedback on gas that did not end up in groups in {\it ZCool\_SF\_SN\_AGN} but did end up in groups in {\it NoCool} (i.e, ejected gas), we present Fig.\ 12.  For each group at $z=0$ we compute the distribution of the logarithm of the ratio of the particle entropy in {\it ZCool\_SF\_SN\_AGN} to the particle entropy in {\it NoCool} for particles that end up either inside or outside of groups in {\it ZCool\_SF\_SN\_AGN}.  This is done on a particle-by-particle basis by exploiting the identical initial conditions and unique particle IDs of gas in the simulations.  For comparison, we also compute the analogous distributions for {\it ZCool\_SF}.  The curves in Fig.\ 12 represent the median histograms for groups in the mass range $13.25 \le \log_{10}(M_{500}/M_\odot) \le 14.25$.

Fig.\ 12 shows that the $z=0$ distribution of entropy magnifications relative to {\it NoCool} for particles that end up {\it inside} groups in {\it ZCool\_SF\_SN\_AGN} is similar to that of particles that end up either inside or outside of groups in {\it ZCool\_SF}, which has no feedback\footnote{Note that the distributions for {\it ZCool\_SF} are not centered on 0 (in log space), as one might naively expect.  As discussed in Section 3.2.1, this is most likely because the gravitational heating experienced by the IGrM in this run exceeds that of the same gas in {\it NoCool} due to the reduced baryon fraction of groups in the former, which allows gas to fall further into the potential wells of groups and therefore to shock to a higher entropy.}.  This can only mean that the gas that ends up constituting the \igm\ in {\it ZCool\_SF\_SN\_AGN} has not been significantly heated by BH feedback, as discussed in Section 3.4.

\begin{figure}
\includegraphics[width=\columnwidth]{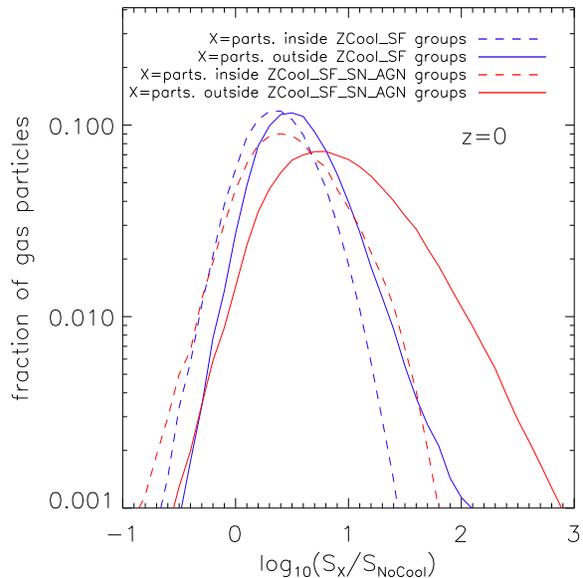}
\caption{The distribution of the ratio of the particle entropy in {\it ZCool\_SF\_SN\_AGN} (red curves) or {\it ZCool\_SF} (blue curves) to the particle entropy in {\it NoCool}, both for particles that end up inside (dashed curves) and outside (solid curves) groups in {\it ZCool\_SF\_SN\_AGN} or {\it ZCool\_SF}.  The similarity of the dashed red and dashed blue curves indicates that AGN feedback has not strongly heated the gas that ends up constituting the IGrM.  However, the solid red curve, which is shifted to higher entropy magnifications, indicates that the gas outside groups in the run with AGN feedback has been strongly heated and was ejected from the system as a result.}
\end{figure}

By contrast, the distribution of entropy magnifications for particles that end up {\it outside} groups in {\it ZCool\_SF\_SN\_AGN} (i.e., ejected gas) is shifted away from the other curves, implying that the gas has been strongly heated on average, and the distribution is skewed to higher ratios (i.e., a larger fraction of the particles have very high entropy ratios).  Typically, the entropy of an ejected gas particle in the run with AGN feedback has had its entropy raised by a factor of $\sim 10$ over that in {\it NoCool}, although many ejected particles receive boosts of up to $\sim 100$ times (and higher).



\subsection{How was the gas heated?}

Was the gas heated abruptly or gradually over time?  Until now we have focused only on {\it median} entropy histories, which is the typical gas particle entropy as a function of time.  The advantage of the median is that it gives an indication of what the population of particles as a whole is doing.  The disadvantage is that it effectively `smooths' out the histories of individual particles.  Thus, the median history is not particularly useful for assessing if the heating process occurs gradually or through one abrupt episode.  Gradual heating could happen, for example, if outflowing gas experiences numerous weak shocks from collisions with hydrostatic/inflowing gas in galaxies or with other outflowing gas (e.g., newly outflowing gas could catch up with gas heated earlier), or if the gas re-collapses and experiences weak accretion shocks.  However, it is difficult to assess the importance of gradual vs.\ abrupt heating based on the median history of a large sample of particles.  A better approach is to look at the thermal history of individual ejected gas particles.

\begin{figure*}
\includegraphics[width=18cm]{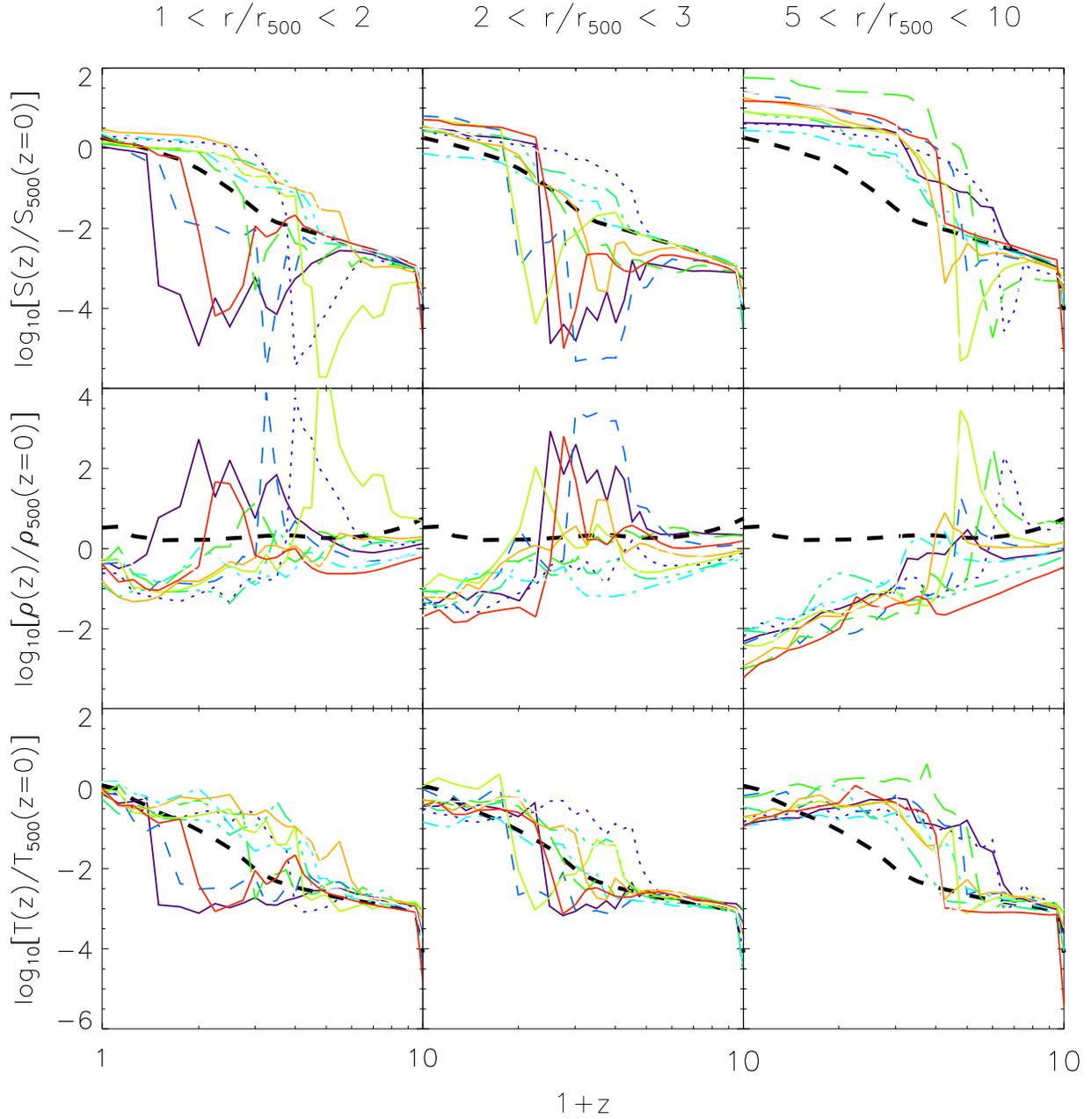}
\caption{The entropy (top), density (middle), and temperature (bottom) histories of randomly-selected ejected gas particles associated with a randomly-selected galaxy group in the {\it ZCool\_SF\_SN\_AGN} run.  Ten ejected gas particles are randomly-selected in 3 radial bins: $1 < r/r_{500} < 2$ (left), $2 < r/r_{500} < 3$ (center), and $5 < r/r_{500} < 10$ (right). The thick dashed black curve represents the median entropy history of gas constituting the present-day \igm\ (i.e., $r < r_{500}$) and the coloured curves represent the 10 randomly-selected ejected gas particles.  Broadly speaking, based on the entropy histories there are two classes of ejected gas particles: those which were heated gradually and those which were heated abruptly.  The latter probably correspond to particles that were directly heated by BHs, while the former are likely those particles which were heated from collisions with the outflowing directly-heated gas.}
\label{fig:trajectories}
\end{figure*}

We randomly select ejected gas particles associated with a randomly-selected galaxy group in {\it ZCool\_SF\_SN\_AGN}.  For reference, the randomly-selected group has a total mass of M$_{500}(z=0) \approx 9.2 \times 10^{13}$ M$_\odot$ with $T_{500}(z=0) \approx 1.3 \times 10^7$ K and $S_{500}(z=0) \approx 186$ keV cm$^2$. (Note that we have also randomly selected other groups for comparison and find very similar results.).  We randomly select 10 particles in several radial bins (the same bins as adopted in the right panel of Fig.\ 11).  In Fig.\ 13 we show the entropy, temperature, and density histories of these particles in three of the radial bins (see caption).

Examining first the entropy histories (top panels) we see that, broadly speaking, the particles may be broken up into two classes: one where the entropy rises gradually, similar to (but in excess of) the gas destined to form the \igm\ (i.e., $r < r_{500}$; thick dashed black curve), and another where the entropy first drops before being abruptly raised to nearly its final value.  

The physical interpretation of the latter population is simple: it represents gas that significantly cooled (radiatively) and collapsed to the centre of its halo before being directly heated by a supermassive BH.  We note that the entropy jump experienced by this gas (often $\sim 4$ orders of magnitude) is about what we would expect given that this gas cooled down to $\sim 10^4$ K (see bottom panels) before being heated\footnote{The temperature histories typically  only show a jump of $\sim 3$ orders of magnitude, rather than 4, which would appear to be inconsistent with the entropy jump.  However, we note that the time scale for a single hot particle to cool adiabatically by free expansion is $\sim h/c_s$, where $h$ is the SPH kernel length $[\sim (m/\rho)^{1/3}]$ (where $m$ and $\rho$ are the gas particle mass and density) and $c_s$ the sound speed. For groups with $c_s$ of a few hundred kilometers per second, this timescale is short compared to the interval between snapshots.  Thus, we expect directly-heated gas particles to cool due to adiabatic expansion.  However, this kind of expansion does not affect the entropy.} by $\Delta T_{\rm heat} = 10^8$ K (see Section 2.1).  Some additional entropy generation occurs after the direct heating episode, which is likely due to weak shocks experienced by the gas as it flows outward.

In order to determine the nature of the heating of the class of particles whose entropy increases much more gradually with time (but still in excess of the gas that ends up in the group), one needs to also examine the density and temperature histories.  For this class of particles, most of the entropy production occurs between $1 \la z \la 4$.  Over this time period the density of these particles is typically slightly decreasing with time towards the present day.  This indicates that the gas is outflowing and, as it does so, is experiencing numerous weak shocks.  In fact, it is likely the gas is outflowing {\it because} it has experienced weak shocks, from collisions with rapidly outflowing directly-heated gas.

That the outflowing, directly-heated gas shock heats other gas leading to further ejection can also be arrived at through other considerations.  In particular, the mass of directly-heated gas ($m_{\rm heat}$) is given by\footnote{This is actually an upper limit to the mass of directly-heated gas, since it neglects BH growth due to mergers and assumes that a gas particle can be heated only once.}

\begin{equation}
m_{\rm heat} = \epsilon_r \epsilon_f m_{\rm BH} c^2 / [(3/2) k_B \Delta T_{\rm heat}] \ \mu m_{\rm H} \ \ \ ,
\end{equation}

\noindent where $\epsilon_r \epsilon_f = 0.015$ is the feedback efficiency (see Booth \& Schaye 2009) and $\Delta T_{\rm heat} = 10^8$ K $\approx 8.6/k_B$ keV is the energy injected per particle (see Section 2.1).

Over the relatively small range in halo masses considered here, we find that the total mass of BHs within $r_{500}$ is $m_{\rm BH}(<r_{500}) \approx 2 \times 10^{-5}$ M$_{500}$.  Filling in the other numbers yields the approximate scaling between the heated gas mass and the group mass:

\begin{equation}
m_{\rm heat} \approx 1.3 \times 10^{-2} M_{500} \nonumber
\end{equation}

One can compare this estimate of the directly-heated mass of gas with the typical mass of ejected gas (see Fig.\ 8), which approximately scales with $M_{500}$ as

\begin{equation}
m_{\rm ejected} \approx f_b (-0.5  \log_{10}[M_{500}/3\times10^{13} M_\odot] + 0.46)  \ \ \ .
\end{equation}

Therefore, at our median mass of $M_{500} \approx 3\times10^{13} M_\odot$ the ratio of ejected to directly-heated gas mass is $\approx 6.2$.  This indicates that the directly-heated gas is able to heat neighbouring low-density gas, leading to additional mass outflow.  But note this process becomes less effective with increasing group mass.  For example, for the randomly-selected group studied in Fig.\ 13, which has a mass of $M_{500} \approx 9.2\times10^{13} M_\odot$, the expected ratio of ejected to directly-heated gas mass is $\approx 2.9$.  This ratio is approximately consistent with the fraction of particles that experience large entropy jumps in Fig.\ 13, confirming our hypothesis that these particles have been directly heated by the AGN.

\subsection{Sensitivity to parameters of the AGN feedback model}

One of the main motivations for the \owls\ project is to quantify the
effects of varying subgrid physics in the simulations. This is done by 
varying the
parameters of a given subgrid physics module, or by switching on/off the
various modules one at a time.  In a future paper we will present a
detailed comparison of groups from all of the $\sim 50$ \owls\ runs,
which, among other things, vary the parameters of the prescriptions for
stellar populations, chemodynamics, radiative cooling, and SN and AGN
feedback.  For the purposes of the present study, however, it is worth
briefly discussing the sensitivity of our results/conclusions to parameter
variations for our implementation of AGN feedback (see Booth \& Schaye
2009 for a detailed description of the model).

As already noted in Section 2.1, a certain fraction of the rest mass
energy of accreted gas, $\epsilon$, is used to heat a number of
randomly-selected neighbouring gas particles. Through systematic 
variation, Booth \& Schaye (2009)
found that a value of $\epsilon \approx0.015$ reproduces the local relations
between BH mass and stellar mass and velocity dispersion, as well as the
$z=0$ cosmic BH density.  We therefore decided to leave $\epsilon$ fixed
at this value.  BHs in the simulation store up this accreted `feedback
energy' until it is sufficient to raise the temperature of $n_{\rm heat}$
gas particles by $\Delta T_{\rm heat}$.  We have explored the effects of
varying these two parameters on the properties of the IGrM.  In
particular, we have tried varying $n_{\rm heat}$ at fixed $\Delta T_{\rm
heat}$ and $\Delta T_{\rm heat}$ at fixed $n_{\rm heat}$, as well as
fixing the product $n_{\rm heat} \Delta T_{\rm heat}$ (which implies a
fixed level of energy injection per feedback episode) while varying both
parameters.

We find that the results are quite insensitive to variations in $n_{\rm
heat}$ at fixed $\Delta T_{\rm heat}$, but are somewhat sensitive to
variations in $\Delta T_{\rm heat}$ at fixed $n_{\rm heat}$.  This is not
unexpected, since if $\Delta T_{\rm heat}$ is set to values that are
comparable to, or smaller than, the temperature of the local ambient hot
medium, the heating will be inefficient.  This is because the gas is
heated at fixed density, implying $\Delta T/T = \Delta S/S$, so that a
small fractional change in temperature corresponds to a small fractional
change in entropy.

In terms of the {\it median} entropy of the IGrM, we find that the result is
insensitive to the exact value of $\Delta T_{\rm heat}$ so long as it is
sufficiently high to expel gas from {\it the low-mass progenitors} of
galaxy groups.  As we have shown above, it is the ejection of gas from
these high-z systems that yields the required truncation of the \igm\
entropy distribution.  To give one specific example, we found very similar
median \igm\ entropies (compared to {\it ZCool\_SF\_SN\_AGN}, which uses 
$\Delta T_{\rm heat} = 10^8$ K and $n_{\rm heat} = 1$) for a run
with $\Delta T_{\rm heat} = 10^7$ K and $n_{\rm heat} = 10$.  Note,
however, that in this case $\Delta T_{\rm heat}$ is comparable to the
virial temperature of the $z=0$ groups we are studying.  The argument
presented above therefore implies that heating will be ineffective in
these haloes today, and indeed we find that the entropy at
the centers of the groups is lower than observed
(even though the median entropies of the groups remain similar to what is
observed).  In other words, decreasing $\Delta T_{\rm heat}$ to this level
results in a pile-up of low entropy gas, since the entropy added to the
gas by AGN is not sufficient for the gas to rise buoyantly out of the
group core.

The upshot of the above is that as long as the AGN is `able to do
something' (i.e., the injected energy is not immediately radiated
away), it tends to yield the same \igm\ properties, independent of the
exact choice for the values of the parameters of the model.

\section{Discussion and Conclusions}


In Paper I we demonstrated that the {\it ZCool\_SF\_SN\_AGN} run that is part of the large \owls\ suite of simulations (Schaye et al.\ 2010) simultaneously reproduces an impressively wide range of X-ray and optical properties of groups, including the entropy, temperature, and metallicity distributions and their dependence on halo mass, as well as the stellar mass of the group and the central brightest galaxy (CBG) and the stellar age of the CBG (see also Sijacki et al.\ 2007; Puchwein et al.\ 2008; Fabjan et al.\ 2010, who also find improved agreement with the data when AGN feedback is included in their simulations).  In the present study we have used a number of additional \owls\ runs (including the effectively non-radiative and no feedback runs {\it NoCool} and {\it ZCool\_SF}, respectively) to deduce exactly how the \igm\ entropy is raised in the {\it ZCool\_SF\_SN\_AGN} run.  

Somewhat surprisingly, we find that none of the previously proposed mechanisms for raising the entropy of the \igm\ are a good description of what happens in the simulation.  The excess entropy of the gas within groups today is {\it not} the result of heating of that gas.  On the contrary, we find that the gas within groups today is that which preferentially avoided strong heating by BH feedback.  Instead, we find that the excess entropy is the result of selectively removing the lowest entropy from the system by star formation and, much more importantly, by the ejection of gas from low-mass progenitor haloes, particularly over the redshift range $z \approx 2-4$ (see Fig.\ 11).  

The idea of selective removal of low entropy material as a mechanism for raising the entropy of the \igm\ goes back to the studies of VB.  Originally, these authors envisaged that the lowest entropy material would end up forming stars [although Voit \& Bryan (2001) and Voit et al.\ (2002) did note the possibility that extreme heating could also remove the lowest entropy gas].  However, it is now clear that in order to reproduce the observed excess entropy via cooling/star formation alone, an unacceptably large fraction of the gas would need to be converted to stars.  Bower et al.\ (2008; hereafter B08) proposed a simple modification of the VB truncation mechanism, in which BH feedback specifically targets the lowest entropy gas.  In this way, the entropy of the \igm\ is still raised by the removal of the lowest entropy gas while allowing the possibility of retaining reasonable stellar mass fractions.  Indeed, B08 implemented this simple modification of the VB mechanism into the Durham {\small GALFORM} semi-analytic model of galaxy formation and showed that it can yield simultaneous matches to the mean thermal properties of the \igm\ and to the stellar content of group galaxies.  

While our detailed cosmological hydrodynamic simulations also show that the primary mechanism for generating excess entropy in groups is through ejection of low entropy gas, the way in which this happens differs from that proposed by B08.  In particular, B08 (see also Bower et al.\ 2006) implemented a prescription for `radio mode' AGN feedback which becomes effective in high mass systems where quasi-hydrostatic gaseous atmospheres can develop.  In their model ejection only occurs if the energy imparted to the gas exceeds the gravitational binding energy of the present-day group.  Consequently, the energy required to eject the gas is substantial (see B08 for further discussion).  By contrast, in our simulation the ejection occurs at high redshift when the low entropy gas is located in the shallower potential wells of the low-mass progenitors of groups.  Consequently, the energy required to eject the gas is lower than that required by the model of B08.  The ejection in our simulations occurs during the phase when the black hole accretion rate is close to Eddington limited.  In other words, we find that it is `quasar mode' AGN feedback (as opposed to `radio mode') that is responsible for raising the entropy of the \igm\ over and above that due to gravitational shock heating.

In the present study we have of course not explored the full range of conceivable AGN feedback models.  The simulations are therefore not guaranteed to capture the physics of AGN feedback correctly.  However, we are encouraged by the fact that the model simultaneously reproduces a wide range range of global and structural X-ray/optical properties of galaxy groups (see Paper I) with virtually no tuning of the model parameters (in terms of AGN feedback, only the efficiency of feedback was tuned so that the simulations reproduce the normalization of the local black hole scaling relations; see Booth \& Schaye 2009).  A direct test of the model will be to search for outflows at high redshift to confirm their existence and validate the model.  Indeed, there are already signs that such `quasar mode' outflows are operating both at low and high redshift (see, e.g., Moe et al.\ 2009; Dunn et al.\ 2010; Tombesi et al.\ 2010a,b; Feruglio et al.\ 2010; Raimundo et al.\ 2010).  Determining how 
common these outflows are and what their typical properties are (i.e., whether they are ejecting sufficient mass on average to explain the overall entropy and baryon mass fractions of local groups) will be key.

How realistic is the current implementation of AGN feedback given our
existing knowledge of real quasar-driven outflows?  Recent observations
suggest outflow velocities up to a significant fraction of the speed of
light on very small scales ($\sim$ pc; e.g., Tombesi et al.\ 2010a). Of course there is no hope of
being able to resolve such tiny scales in cosmological simulations at
present.  However, the hot, outflowing gas will expand and cool
adiabatically and as long as: i) the simulations capture the outflow
when the velocities are still supersonic with respect to the the IGrM;
and, ii) the cooling time of the shocked gas is large; then the
simulations should retrieve approximately the same result for a fixed
amount of energy. In this respect, an important parameter of the model
is $\Delta T_{\rm heat}$.  If $\Delta T_{\rm heat} \la T_{\rm vir}$,
then heating will be inefficient\footnote{As we discussed in Section 4.4, in terms of
reproducing the median entropies of the IGrM, what is impor-
tant, is the depth of the potential well (characterised by Tvir) of the
progenitors of groups, rather than that of present-day
groups.} (i.e., it violates the second
condition), which changes the physics and the results will be sensitive
to the exact value $\Delta T_{\rm heat}$ (see Section 4.4 for further
discussion).  This is the reason why we have adopted a `high' value for
$\Delta T_{\rm heat}$ - it guarantees efficient feedback on the scale of
groups (but note that this choice in no way guarantees that the
simulations will reproduce the observations).

Another possible way to test the model may be through observations of
cosmic rays (CRs), since the AGN feedback results in some very strong   
shocks, with Mach numbers approaching $\sim 100$ (see Fig.~13).  This
may result in the production of CRs, but whether or not this will be
easily distinguishable from CRs produced in shocks driven by SNe and   
mergers (e.g., Pfrommer et al.\ 2007) will require further detailed 
modelling.

It is worth briefly commenting on the suitability of the SPH formalism
to the problem at hand. Since SPH by its nature will smooth out shocks,
can we be confident that the rate of shock heating in the simulations is
correct?  We have verified that a version of the AGN model run with
8 times lower mass resolution (2 times lower spatial resolution) yields remarkably similar entropy distributions for the IGrM.  Also, as long as the radiative cooling time in the post-shock gas is long, the correct result should be obtained, since the energy is
conserved and the gas will expand in both cases (i.e., in a real shock 
and in a SPH shock). The implementation of AGN feedback in our
simulations guarantees that the post-shock gas has a long cooling time.

It is interesting to ask the question whether only AGN feedback is
capable
of `raising' the entropy of the \igm\ to the observed level.  For example,
perhaps an alternative implementation of SN feedback in the simulations
could achieve a similar result.  We find that it is difficult to rule 
out this
possibility based on energy arguments alone.  We have calculated the
total energy injected by BHs within $r_{500}$ in our simulation as
$[\epsilon_r \epsilon_f / (1 - \epsilon_r)] m_{\rm BH}(< r_{500}) c^2$,
where $\epsilon_r$ and $\epsilon_f$ are the radiative efficiency of the BH
accretion disk and the fraction of emitted energy assumed to couple to the
gas and $m_{\rm BH}(< r_{500})$ is the total mass of BHs within $r_{500}$.
This yields a typical total injection energy of a few $10^{61}$ ergs, or
approximately 1-2 keV per particle, assuming $M_{\rm baryon}(r_{500}) =
f_b M_{500}$.  The maximum amount of energy available from SNe is
estimated as follows.  We sum the {\it initial} (i.e., ignoring stellar mass loss) stellar mass of star
particles within $r_{500}$ in {\it ZCool\_SF\_SN\_AGN} (which reproduces
the $z=0$ stellar mass fractions of observed groups, see Paper I).
Assuming a Chabrier IMF, we calculate the number of stars in the mass
range $8-100 M_\odot$, all of which are assumed to yield core-collapse
SNe.  Finally, the energy per SN is assumed to be $10^{51}$ ergs. 
Typically, we find that
the energy available from SNe is $0.5-1.0$ times that injected by BHs in
our simulation (note, however, that the {\it maximum} amount of energy
that can be injected by BHs is an order of magnitude larger, but that the
adopted BH efficiency was tuned to match the normalisation of the $z=0$ 
BH scaling relations -
see Booth \& Schaye 2009).  Similar results are obtained if we use
$r_{200}$ as our limiting radius, rather than $r_{500}$.

Energetically, therefore, the contribution from SNe is significant.
However, whether or not SNe can significantly raise the entropy of the \igm\
depends on the importance of radiative losses.  Since the energy injected by
SNe will likely be much more spatially-distributed than that by BHs, one
expects radiative losses to be more important for the former.  Indeed,
this is the reason why {\it ZCool\_SF\_SN} yields much higher stellar mass
fractions and lower \igm\ entropies than {\it ZCool\_SF\_SN\_AGN} in the
present study.  We also note that {\it none} of the many variations of the
SN-driven feedback model parameters explored as part of the \owls\ project
yield group properties that are similar to {\it ZCool\_SF\_SN\_AGN}
(this will be presented in detail in a future paper).  A wide range of
implementations of SN-driven feedback have been explored in previous
studies and similar conclusions have been drawn (e.g., Thacker \& Couchman
2000; Borgani et al.\ 2002, 2004; Kay et al.\ 2004).

Two notable exceptions are Kay (2004) and Dav{\'e} et al.\ (2008).
Dav{\'e} et al.\ (2008) implemented a stellar feedback model in which 
the initial wind mass loading and velocity scale with halo properties as 
expected on larger scales for outflows driven by radiation pressure from 
massive stars on dust grains.  They found groups with stellar mass fractions
that are similar to the observed mass fractions.  It is worth noting, 
however, that the normalisation of their feedback prescription was 
treated as a free parameter and that the amounts of
energy and momentum injected in their outflows both far exceed those 
available from star formation.  Kay (2004) showed that if the energy 
injected by SNe
is highly-targeted (i.e., deposited in a very small amount of gas) and raises
the entropy of the gas to a very high level (he adopted 1000 keV cm$^2$), it
can yield group properties in reasonable agreement with observations.
This result is interesting because in this case the SNe `behave' much like
AGN.  The downside of this approach is that one must fine-tune the
parameters of the SN feedback model to achieve a result that comes about
naturally from a model that includes BH growth and AGN feedback.

In the present study we have concerned ourselves with the establishment of only the global (median) properties of the \igm\ and found that quasar mode AGN feedback is the dominant mechanism for raising the overall entropy of the gas.  However, this statement does not imply that SN feedback or radio mode AGN feedback are unimportant.  As we have already discussed in Section 3.3, SN feedback prevents star formation in high-z galaxies.  This gas ends up as part of the \igm\ but it has not been heated to such an extent that it modifies the {\it median} entropy of the hot gas.  Low accretion rate (i.e., radio mode) AGN feedback is likely important at late times to prevent the development of cooling flows. But it should be kept in mind that by mass the gas within the cooling radius (where the radio mode heating tends to be targeted) represents only about 10-20\% of the total.  It is therefore difficult for radio mode feedback to affect the {\it global} properties of the gas, which is necessary in order to match observations of groups [see also the recent study of Fontanot et al.\ (2010) who find that semi-analytic models that invoke radio mode only, and whose parameters have been adjusted to reproduce the galaxy population, tend to greatly overproduce the observed radio emission].

Finally, we have focused on the global thermal properties of the hot gas in groups and low-mass clusters.  However, more massive galaxy clusters also show evidence for excess entropy, at least in the central regions (see, e.g., Pratt et al.\ 2010).  At present, it is widely believed that this excess entropy is due to AGN radio mode feedback [in `cool core' (CC) clusters] or cluster-cluster mergers [in `non-cool core' (NCC) clusters].  However, there are problems with both of these explanations.  In the case of radio mode feedback in CC clusters, while there may be sufficient energy to offset radiative cooling losses (e.g., Dunn \& Fabian 2008), there does not appear to be sufficient energy to explain how these systems obtained their high entropies in the first place if the heating occurred via radio mode after cluster formation (McCarthy et al.\ 2008; NCC clusters cannot be explained in this way either).  As for mergers being an explanation for NCC clusters, Poole et al.\ (2006, 2008) used high-resolution SPH simulations of cluster mergers to show that the dense, low entropy gas in the cores of clusters is generally robust to mergers (although see Mitchell et al.\ 2009 and ZuHone 2010 who argued that SPH simulations may underestimate the degree of mixing during mergers).  In any case, the physical origin of the high entropy in clusters (and its scatter) is still very much an open question.  

AGN feedback at high redshift (in quasar mode) may provide an alternative explanation for the excess entropy of the ICM in both CC and NCC clusters.  Naively one might rule the quasar mode hypothesis out by noting that BHs are not sufficiently energetic to eject gas from the most massive systems (which our simulations confirm; see Fig.\ 3).  However, we note that, unlike groups, the excess entropy in clusters is confined to relatively small radii (the gas at intermediate/large radii is described well by the self-similar model).  Thus, to explain the excess entropy in clusters the gas may not have to be ejected to beyond the virial radius.  Heating it at high redshift so that it ends up at intermediate/large radii may be sufficient.  We intend to present the implications of our model for massive galaxy clusters in a future study.

\section*{Acknowledgments}

The authors would like to thank the anonymous referee for suggestions that improved the paper.  They also thank thank Michael Balogh, Mark Voit, Daisuke Nagai, Pierluigi Monaco, Eugene Churazov, Ming Sun, Franco Vazza, and Stefania Giodini for helpful comments and suggestion and Marijn Franx for the suggestion to turn off the AGN at varying redshift.
IGM acknowledges support from a 
Kavli Institute Fellowship at the University of Cambridge.  The simulations presented here were run on Stella, the LOFAR
BlueGene/L system in Groningen, on the Cosmology Machine at the
Institute for Computational Cosmology in Durham as part of the Virgo
Consortium research programme, and on Darwin in Cambridge. This work
was sponsored by National Computing Facilities Foundation (NCF) for
the use of supercomputer facilities, with financial support from the
Netherlands Organization for Scientific Research (NWO). 
This work was supported by an NWO VIDI grant.

\section*{Appendix: Effects of UV/X-ray photoheating on the thermal history of galaxy groups}

All of the \owls\ runs include photoheating from the Haardt \& Madau (2001) model for the UV/X-ray background from galaxies and quasars.  In order to assess the effects of radiative cooling, star formation, and feedback on the \igm\ we therefore also included photoheating in our baseline {\it NoCool} model.  Here we compare the {\it NoCool} model to a non-radiative simulation which omits UV/X-ray photoheating.

Fig.\ 14 compares the thermal histories of the {\it NoCool} and non-radiative runs.  While it is clearly important at early times ($z \ga 2$), photoheating has no important consequences for the thermodynamic properties of the present-day IGrM.   

\begin{figure}
\includegraphics[width=\columnwidth]{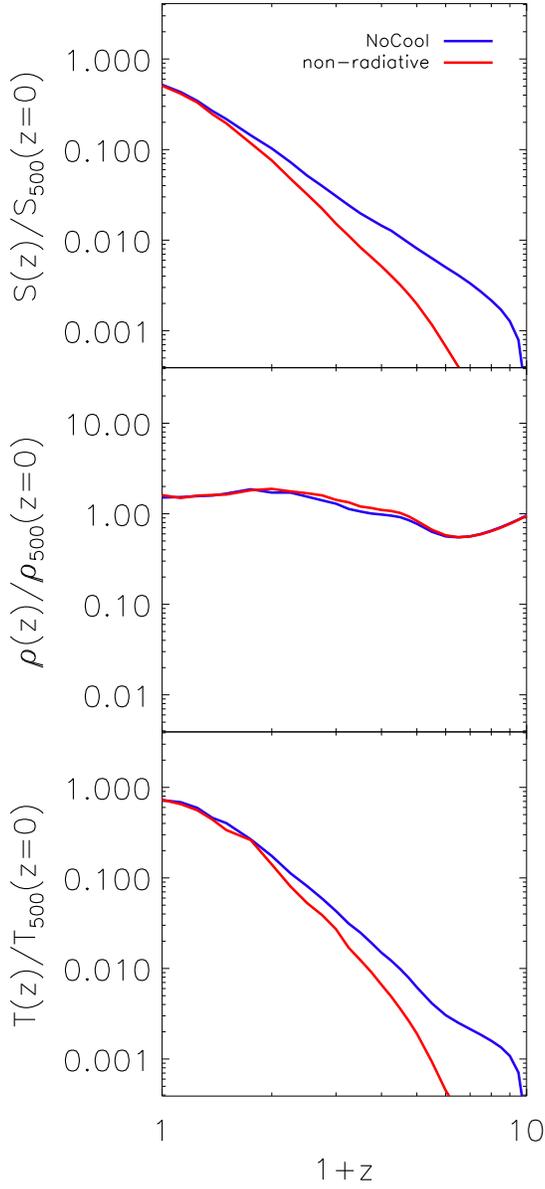}
\caption{Comparison of the thermal histories of the {\it NoCool} run which, like all other runs investigated in this study includes UV/X-ray photoheating, and a non-radiative run, which neglects photoheating.  The curves have the same meaning as in Fig.\ 5.  UV/X-ray photoheating is unimportant at late times, after group formation.}
\label{fig:thermal_history_nonrad}
\end{figure}

\bsp

\label{lastpage}


\begin{thebibliography}{99}

\bibitem[Babul et al.(2002)]{2002MNRAS.330..329B} Babul, A., Balogh, M.~L., 
Lewis, G.~F., \& Poole, G.~B.\ 2002, \mnras, 330, 329 

\bibitem[Balogh et al.(1999)]{1999MNRAS.307..463B} Balogh, M.~L., Babul, 
A., \& Patton, D.~R.\ 1999, \mnras, 307, 463 

\bibitem[Balogh et al.(2001)]{2001MNRAS.326.1228B} Balogh, M.~L., Pearce, 
F.~R., Bower, R.~G., \& Kay, S.~T.\ 2001, \mnras, 326, 1228 

\bibitem[Balogh et al.(2008)]{2008MNRAS.385.1003B} Balogh, M.~L., McCarthy, 
I.~G., Bower, R.~G., \& Eke, V.~R.\ 2008, \mnras, 385, 1003 

\bibitem[Booth 
\& Schaye(2009)]{2009MNRAS.398...53B} Booth, C.~M., \& Schaye, J.\ 2009, \mnras, 398, 53 

\bibitem[Borgani et al.(2002)]{2002MNRAS.336..409B} Borgani, S., Governato, 
F., Wadsley, J., Menci, N., Tozzi, P., Quinn, T., Stadel, J., 
\& Lake, G.\ 2002, \mnras, 336, 409 

\bibitem[Borgani et al.(2004)]{2004MNRAS.348.1078B} Borgani, S., et al.\ 
2004, \mnras, 348, 1078 

\bibitem[Bower(1997)]{1997MNRAS.288..355B} Bower, R.~G.\ 1997, \mnras, 288, 
355 

\bibitem[Bower et al.(2006)]{2006MNRAS.370..645B} Bower, R.~G., Benson, 
A.~J., Malbon, R., Helly, J.~C., Frenk, C.~S., Baugh, C.~M., Cole, S., 
\& Lacey, C.~G.\ 2006, \mnras, 370, 645 

\bibitem[Bower et al.(2008)]{2008MNRAS.390.1399B} Bower, R.~G., McCarthy, 
I.~G., \& Benson, A.~J.\ 2008, \mnras, 390, 1399 

\bibitem[Boyle 
\& Terlevich(1998)]{1998MNRAS.293L..49B} Boyle, B.~J., \& Terlevich, R.~J.\ 1998, \mnras, 293, L49 

\bibitem[Bryan(2000)]{2000ApJ...544L...1B} Bryan, G.~L.\ 2000, \apjl, 544, 
L1 

\bibitem[Chabrier(2003)]{2003PASP..115..763C} Chabrier, G.\ 2003, \pasp, 
115, 763 

\bibitem[Crain et al.(2007)]{2007MNRAS.377...41C} Crain, R.~A., Eke, V.~R., 
Frenk, C.~S., Jenkins, A., McCarthy, I.~G., Navarro, J.~F., 
\& Pearce, F.~R.\ 2007, \mnras, 377, 41 

\bibitem[Crain et al.(2009)]{2009MNRAS.399.1773C} Crain, R.~A., et al.\ 
2009, \mnras, 399, 1773 

\bibitem[Dalla Vecchia 
\& Schaye(2008)]{2008MNRAS.387.1431D} Dalla Vecchia, C., \& Schaye, J.\ 2008, \mnras, 387, 1431 

\bibitem[Dav{\'e} et al.(2002)]{2002ApJ...579...23D} Dav{\'e}, R., Katz, 
N., \& Weinberg, D.~H.\ 2002, \apj, 579, 23 

\bibitem[Dav{\'e} et al.(2008)]{2008MNRAS.391..110D} Dav{\'e}, R., 
Oppenheimer, B.~D., \& Sivanandam, S.\ 2008, \mnras, 391, 110 

\bibitem[Duffy et al.(2010)]{2010MNRAS.405.2161D} Duffy, A.~R., Schaye, J., 
Kay, S.~T., Dalla Vecchia, C., Battye, R.~A., 
\& Booth, C.~M.\ 2010, \mnras, 405, 2161 

\bibitem[Dunn et al.(2010)]{2010ApJ...709..611D} Dunn, J.~P., et al.\ 2010, 
\apj, 709, 611 

\bibitem[Dunn 
\& Fabian(2008)]{2008MNRAS.385..757D} Dunn, R.~J.~H., \& Fabian, A.~C.\ 2008, \mnras, 385, 757 

\bibitem[Eke et al.(1998)]{1998ApJ...503..569E} Eke, V.~R., Navarro, J.~F., 
\& Frenk, C.~S.\ 1998, \apj, 503, 569 

\bibitem[Evrard 
\& Henry(1991)]{1991ApJ...383...95E} Evrard, A.~E., \& Henry, J.~P.\ 1991, \apj, 383, 95 

\bibitem[Fabjan et al.(2010)]{2010MNRAS.401.1670F} Fabjan, D., Borgani, S., 
Tornatore, L., Saro, A., Murante, G., \& Dolag, K.\ 2010, \mnras, 401, 1670 

\bibitem[Ferland et al.(1998)]{1998PASP..110..761F} Ferland, G.~J., 
Korista, K.~T., Verner, D.~A., Ferguson, J.~W., Kingdon, J.~B., 
\& Verner, E.~M.\ 1998, \pasp, 110, 761 

\bibitem[Feruglio et al.(2010)]{2010arXiv1006.1655F} Feruglio, C., 
Maiolino, R., Piconcelli, E., Menci, N., Aussel, H., Lamastra, A., 
\& Fiore, F.\ 2010, \aa, submitted (arXiv:1006.1655)

\bibitem[Fontanot et al.(2010)]{2010arXiv1006.5717F} Fontanot, F., 
Pasquali, A., De Lucia, G., van den Bosch, F.~C., Somerville, R.~S., 
\& Kang, X.\ 2010, \mnras, submitted (arXiv:1006.5717)

\bibitem[Frenk et al.(1999)]{1999ApJ...525..554F} Frenk, C.~S., et al.\ 
1999, \apj, 525, 554 

\bibitem[Finoguenov et al.(2002)]{2002ApJ...578...74F} Finoguenov, A., 
Jones, C., B{\"o}hringer, H., \& Ponman, T.~J.\ 2002, \apj, 578, 74 

\bibitem[Giodini et al.(2009)]{2009ApJ...703..982G} Giodini, S., et al.\ 
2009, \apj, 703, 982 

\bibitem[Giodini et al.(2010)]{2010ApJ...714..218G} Giodini, S., et al.\ 
2010, \apj, 714, 218 

\bibitem[Gonzalez et al.(2007)]{2007ApJ...666..147G} Gonzalez, A.~H., 
Zaritsky, D., \& Zabludoff, A.~I.\ 2007, \apj, 666, 147 

\bibitem[Haardt 
\& Madau(2001)]{2001cghr.confE..64H} Haardt, F., \& Madau, P.\ 2001, Clusters of Galaxies and the High Redshift Universe Observed in X-rays, XXXVIth Rencontres de Moriond , XXIst Moriond Astrophysics Meeting, March 10-17, 2001 Savoie, France. Edited by D.M. Neumann \& J.T.T. Van (arXiv:astro-ph/0106018)
  
\bibitem[H{\"a}ring 
\& Rix(2004)]{2004ApJ...604L..89H} H{\"a}ring, N., \& Rix, H.-W.\ 2004, \apjl, 604, L89 

\bibitem[Kaiser(1991)]{1991ApJ...383..104K} Kaiser, N.\ 1991, \apj, 383, 
104 

\bibitem[Kauffmann et al.(2003)]{2003MNRAS.346.1055K} Kauffmann, G., et 
al.\ 2003, \mnras, 346, 1055 

\bibitem[Kay et al.(2004)]{2004MNRAS.355.1091K} Kay, S.~T., Thomas, P.~A., 
Jenkins, A., \& Pearce, F.~R.\ 2004, \mnras, 355, 1091 

\bibitem[Kay(2004)]{2004MNRAS.347L..13K} Kay, S.~T.\ 2004, \mnras, 347, L13 

\bibitem[Knight 
\& Ponman(1997)]{1997MNRAS.289..955K} Knight, P.~A., \& Ponman, T.~J.\ 1997, \mnras, 289, 955 

\bibitem[Kravtsov et al.(2005)]{2005ApJ...625..588K} Kravtsov, A.~V., 
Nagai, D., \& Vikhlinin, A.~A.\ 2005, \apj, 625, 588 

\bibitem[Johnson et al.(2009)]{2009MNRAS.395.1287J} Johnson, R., Ponman, 
T.~J., \& Finoguenov, A.\ 2009, \mnras, 395, 1287 

\bibitem[Lin 
\& Mohr(2004)]{2004ApJ...617..879L} Lin, Y.-T., \& Mohr, J.~J.\ 2004, \apj, 617, 879 

\bibitem[McCarthy et al.(2007)]{2007MNRAS.376..497M} McCarthy, I.~G., et 
al.\ 2007, \mnras, 376, 497 

\bibitem[McCarthy et al.(2008)]{2008MNRAS.386.1309M} McCarthy, I.~G., 
Babul, A., Bower, R.~G., \& Balogh, M.~L.\ 2008, \mnras, 386, 1309 

\bibitem[McCarthy et al.(2010)]{2010MNRAS.406..822M} McCarthy, I.~G., et 
al.\ 2010, \mnras, 406, 822 

\bibitem[McGee 
\& Balogh(2010)]{2010MNRAS.403L..79M} McGee, S.~L., \& Balogh, M.~L.\ 2010, \mnras, 403, L79 

\bibitem[McNamara et al.(2005)]{2005Natur.433...45M} McNamara, B.~R., 
Nulsen, P.~E.~J., Wise, M.~W., Rafferty, D.~A., Carilli, C., Sarazin, 
C.~L., \& Blanton, E.~L.\ 2005, \nat, 433, 45

\bibitem[Mitchell et al.(2009)]{2009MNRAS.395..180M} Mitchell, N.~L., 
McCarthy, I.~G., Bower, R.~G., Theuns, T., 
\& Crain, R.~A.\ 2009, \mnras, 395, 180 

\bibitem[Moe et al.(2009)]{2009ApJ...706..525M} Moe, M., Arav, N., 
Bautista, M.~A., \& Korista, K.~T.\ 2009, \apj, 706, 525 


\bibitem[Muanwong et al.(2002)]{2002MNRAS.336..527M} Muanwong, O., Thomas, 
P.~A., Kay, S.~T., \& Pearce, F.~R.\ 2002, \mnras, 336, 527 

\bibitem[Mulchaey(2000)]{2000ARA&A..38..289M} Mulchaey, J.~S.\ 2000, \araa, 38, 289 

\bibitem[Nagai 
\& Kravtsov(2004)]{2004ogci.conf..296N} Nagai, D., \& Kravtsov, A.~V.\ 2004, IAU Colloq.~195: Outskirts of Galaxy Clusters: Intense Life in the Suburbs, 296 (astro-ph/0404350)


\bibitem[Pfrommer et al.(2007)]{2007MNRAS.378..385P} Pfrommer, C., 
En{\ss}lin, T.~A., Springel, V., Jubelgas, M., 
\& Dolag, K.\ 2007, \mnras, 378, 385 

\bibitem[Ponman et al.(1999)]{1999Natur.397..135P} Ponman, T.~J., Cannon, 
D.~B., \& Navarro, J.~F.\ 1999, \nat, 397, 135 

\bibitem[Ponman et al.(2003)]{2003MNRAS.343..331P} Ponman, T.~J., 
Sanderson, A.~J.~R., \& Finoguenov, A.\ 2003, \mnras, 343, 331 

\bibitem[Poole et al.(2006)]{2006MNRAS.373..881P} Poole, G.~B., Fardal, 
M.~A., Babul, A., McCarthy, I.~G., Quinn, T., 
\& Wadsley, J.\ 2006, \mnras, 373, 881 

\bibitem[Poole et al.(2008)]{2008MNRAS.391.1163P} Poole, G.~B., Babul, A., 
McCarthy, I.~G., Sanderson, A.~J.~R., 
\& Fardal, M.~A.\ 2008, \mnras, 391, 1163 

\bibitem[Pratt et 
al.(2010)]{2010A&A...511A..85P} Pratt, G.~W., et al.\ 2010, \aap, 511, A85 

\bibitem[Puchwein et al.(2008)]{2008ApJ...687L..53P} Puchwein, E., Sijacki, 
D., \& Springel, V.\ 2008, \apjl, 687, L53 

\bibitem[Puchwein et al.(2010)]{2010MNRAS.tmp..788P} Puchwein, E., 
Springel, V., Sijacki, D., \& Dolag, K.\ 2010, \mnras, in press (arXiv:1001.3018) 

\bibitem[Raimundo et al.(2010)]{2010arXiv1006.4436R} Raimundo, S.~I., 
Fabian, A.~C., Bauer, F.~E., Alexander, D.~M., Brandt, W.~N., Luo, B., 
Vasudevan, R.~V., \& Xue, Y.~Q.\ 2010, \mnras, in press (arXiv:1006.4436)

\bibitem[Schaye 
\& Dalla Vecchia(2008)]{2008MNRAS.383.1210S} Schaye, J., \& Dalla Vecchia, C.\ 2008, \mnras, 383, 1210 

\bibitem[Schaye et al.(2010)]{2010MNRAS.402.1536S} Schaye, J., et al.\ 
2010, \mnras, 402, 1536 

\bibitem[Sijacki et al.(2007)]{2007MNRAS.380..877S} Sijacki, D., Springel, 
V., Di Matteo, T., \& Hernquist, L.\ 2007, \mnras, 380, 877 

\bibitem[Spergel et al.(2007)]{2007ApJS..170..377S} Spergel, D.~N., et al.\ 
2007, \apjs, 170, 377 

\bibitem[Springel(2005)]{2005MNRAS.364.1105S} Springel, V.\ 2005, \mnras, 
364, 1105 

\bibitem[Springel et al.(2005)]{2005MNRAS.361..776S} Springel, V., Di 
Matteo, T., \& Hernquist, L.\ 2005, \mnras, 361, 776 

\bibitem[Sun et al.(2009)]{2009ApJ...693.1142S} Sun, M., Voit, G.~M., 
Donahue, M., Jones, C., Forman, W., \& Vikhlinin, A.\ 2009, \apj, 693, 1142 

\bibitem[Thacker 
\& Couchman(2000)]{2000ApJ...545..728T} Thacker, R.~J., \& Couchman, H.~M.~P.\ 2000, \apj, 545, 728 

\bibitem[Tombesi et 
al.(2010a)]{2010A&A...521A..57T} Tombesi, F., Cappi, M., Reeves, J.~N., Palumbo, G.~G.~C., Yaqoob, T., Braito, V., \& Dadina, M.\ 2010, \aap, 521, A57 

\bibitem[Tombesi et al.(2010b)]{2010ApJ...719..700T} Tombesi, F., Sambruna, 
R.~M., Reeves, J.~N., Braito, V., Ballo, L., Gofford, J., Cappi, M., 
\& Mushotzky, R.~F.\ 2010, \apj, 719, 700 

\bibitem[Tozzi 
\& Norman(2001)]{2001ApJ...546...63T} Tozzi, P., \& Norman, C.\ 2001, \apj, 546, 63 

\bibitem[Vazza(2010)]{2010arXiv1008.0191V} Vazza, F.\ 2010, \mnras, in press (arXiv:1008.0191 )

\bibitem[Voit(2005)]{2005RvMP...77..207V} Voit, G.~M.\ 2005, Reviews of 
Modern Physics, 77, 207 

\bibitem[Voit
\& Bryan(2001)]{2001Natur.414..425V} Voit, G.~M., \& Bryan, G.~L.\ 2001, \nat, 414, 425

\bibitem[Voit et al.(2002)]{2002ApJ...576..601V} Voit, G.~M., Bryan, G.~L., 
Balogh, M.~L., \& Bower, R.~G.\ 2002, \apj, 576, 601 

\bibitem[Voit et al.(2003)]{2003ApJ...593..272V} Voit, G.~M., Balogh, 
M.~L., Bower, R.~G., Lacey, C.~G., \& Bryan, G.~L.\ 2003, \apj, 593, 272 

\bibitem[Voit et al.(2005)]{2005MNRAS.364..909V} Voit, G.~M., Kay, S.~T., 
\& Bryan, G.~L.\ 2005, \mnras, 364, 909 

\bibitem[Wiersma et al.(2009)]{2009MNRAS.393...99W} Wiersma, R.~P.~C., 
Schaye, J., \& Smith, B.~D.\ 2009a, \mnras, 393, 99 

\bibitem[Wiersma et al.(2009)]{2009MNRAS.399..574W} Wiersma, R.~P.~C., 
Schaye, J., Theuns, T., Dalla Vecchia, C., 
\& Tornatore, L.\ 2009b, \mnras, 399, 574 

\bibitem[ZuHone(2010)]{2010arXiv1004.3820Z} ZuHone, J.\ 2010, \apj, submitted
(arXiv:1004.3820)

\end{thebibliography}
\end{document}